\documentclass[APA,STIX1COL]{WileyNJD-v2}

\usepackage{graphicx} 
\usepackage{hyperref}
\hypersetup{
	citecolor= blue, 
	colorlinks=true,
	linkcolor=cyan, 
	filecolor=magenta,      
	urlcolor=cyan,
	pdftitle={Overleaf Example},
	pdfpagemode=FullScreen,
}
\usepackage{natbib}
\usepackage{amsmath}
\usepackage{subfig}
\usepackage{xcolor}
\usepackage{ulem}
\usepackage{amsfonts}
\usepackage{booktabs}
\usepackage{float}

\articletype{Article Type}%

\begin{document}
\title{Detecting changes in space-varying parameters of local Poisson point processes}

\author[1]{Nicoletta D'Angelo}

\authormark{D'ANGELO}

\address[1]{\orgdiv{Department of Economics, Business and Statistics}, \orgname{University of Palermo}, \orgaddress{\state{Italy}, \country{Palermo}}}

\corres{Nicoletta D'Angelo\\ \email{nicoletta.dangelo@unipa.it}}

\abstract[Summary]{
Recent advances in local models for point processes have highlighted the need for flexible methodologies to account for the spatial heterogeneity of external covariates influencing process intensity. In this work, we introduce \textit{tessellated spatial regression}, a novel framework that extends segmented regression models to spatial point processes, with the aim of detecting abrupt changes in the effect of external covariates onto the process intensity.

Our approach consists of two main steps. First, we apply a spatial segmentation algorithm to geographically weighted regression estimates, generating different tessellations that partition the study area into regions where model parameters can be assumed constant. Next, we fit log-linear Poisson models in which covariates interact with the tessellations, enabling region-specific parameter estimation and classical inferential procedures, such as hypothesis testing on regression coefficients.

Unlike geographically weighted regression, our approach allows for discrete changes in regression coefficients, making it possible to capture abrupt spatial variations in the effect of real-valued spatial covariates. Furthermore, the method naturally addresses the problem of locating and quantifying the number of detected spatial changes.

We validate our methodology through simulation studies and applications to two examples where a model with region-wise parameters seems appropriate and to an environmental dataset of earthquake occurrences in Greece.
}


\jnlcitation{\cname{%
\author{N. D'Angelo}} (\cyear{2025}), 
\ctitle{Detecting changes in space-varying parameters of local Poisson point processes}, \cjournal{Environmetrics}.}

\maketitle

\section{Introduction}\label{sec:intro}

Interest in methods for analysing spatial point processes has increased across many fields of science, notably in ecology, epidemiology, geoscience, astronomy, econometrics, and crime research \citep{baddeley:rubak:tuner:15,diggle:13}. When the structure of a given point pattern is observed, it is assumed as a realisation of an underlying generating process, whose properties are estimated and then used to describe the structure of the observed pattern. 
Analysing a spatial point process, the first step is to learn about its first-order characteristics, studying the relationship of the points with the underlying environmental variables that describe the observed heterogeneity. 
The aim is typically to learn about the mechanism that generates these events \citep{moller:98,diggle:13,illian:penttinen:stoyan:stoyan:08}.

The most common point process model taking into account external covariates is the log-linear Poisson point process model, typically globally defined with the process properties and estimated parameters assumed to be constant in all the study area. However, a model with constant parameters, may not adequately represent detailed local variations in the data since the pattern may present spatial variation in the influence of covariates, in the scale or spacing between points, and the abundance of points.
Indeed,  a different way of analysing a point pattern is commonly based on local techniques identifying specific and undiscovered local structure, for instance, sub-regions characterised by different interactions among points, intensity and influence of covariates. 
Throughout the paper we shall distinguish between `global' models, in which the parameters are assumed constant as in regular regression models, and `local' models, in which the parameters are allowed to vary with location.

For spatial point processes, \cite{baddeley:2017local} presented a general \textit{geographically weighted regression} framework based on the local composite likelihood to detect and model gradual spatial variation in any parameter of a spatial point process model (such as Poisson, Gibbs, and Cox processes). In particular, the parameters in the model that govern the intensity, the dependence of the intensity on the covariates, and the spatial interaction between points, are estimated locally. 
In the context of spatial and spatio-temporal point process modeling, the necessity of capturing local variations in the effects of the model's parameters has led to the development of many advanced methodologies that allow for spatially varying parameters \citep{siino2017spatial,siino2018multiscale,d2022locally,d2025non,d2024minimum,d2024stopp,tarantino2024modelling}. In this paper, we only deal with spatial Poisson point processes. For these processes, local likelihood is well-developed, because of its close relationship to probability density estimation from i.i.d. observations \citep{hjort1996locally,loader1996local,loader1999bandwidth}. Geographically weighted logistic regression for pixel presence–absence data \citep{osborne2007non} is equivalent to maximum local likelihood for Poisson processes where the intensity is a loglinear function of the regression parameters (see also \cite{fan2009local,monir2006cluster,ogata1988likelihood}).

However, these local approaches often rely on non-parametric techniques, which can complicate interpretation and standard statistical inference. In this work, we propose a novel method for formalising and fitting a \textit{tessellated} spatial log-linear Poisson point process model, specifically designed to provide clear-cut regions where covariate effects change while maintaining a fixed-effects modelling framework. Indeed, a tessellation is a division of a window into non-overlapping regions, called \textit{tiles}. The main idea behind our proposal is to define a tessellation of the study area that partitions the space into distinct regions, each characterised by a different set of parameter estimates. 
Unlike geographically weighted regression, our approach allows for discrete changes in regression coefficients, making it possible to capture abrupt spatial variations in the effect of real-valued spatial covariates. Therefore, it is advisable to employ tessellated regression when the interest is not on estimating smoothly varying spatial effects but rather identifying and quantifying abrupt changes in model parameters. The main objective of \textit{tessellated regression} is, therefore, the identification of these abrupt changes, represented by tessellations of the analysed region. These delimitations between areas where the effect of a variable changes in explaining the process intensity may be informative of the underlying process generating the events. Indeed, region-wise parameters, may be the effect of unobserved characteristics not taken into account in the fitted model. Moreover, our approach has the advantage of detecting and modeling abrupt spatial variation in covariates' effects within the formal likelihood framework, providing region-wise parameter estimates, confidence intervals, and hypothesis tests.
This makes it particularly useful for applications requiring localised estimation while preserving interpretability within the classical Poisson log-linear modeling framework.
Indeed, the motivation of this research came from
the previous work of 
\cite{dangelo2021locall}, where the authors fitted a local model to earthquake data, accounting for external geological information. That work proved the presence of a clear separation of the analysed region onto subregions where the model parameters completely change. The delimitation of such regions however, was not possible employing the standard local models, as well as the straightforward assessment of the \lq \lq locality'' of a specific covariate.

Our proposed strategy can be seen as the spatial point processes extension of the most known segmented regression model \citep{lerman1980fitting}, commonly employed in time series or, in the general, when a covariate effect is assumed to affect the response variable in a piece-wise manner.
The main advantage of these models lies in the results' interpretability while also achieving a good trade-off with flexibility, typically achieved by non-parametric approaches. Our work represents the first attempt in this direction for point process models. Indeed, in the realm of spatial statistics, \cite{hollaway2024detection} recently presented a spatio-temporal changepoint method that utilises a generalised additive model (GAM) dependent on the 2D
spatial location and the observation time to account for the underlying spatio-temporal geostatistical process. However, to the best of our knowledge, no methods have been yet proposed to study the changes in the effect of covariates in point processes.
For instance, \cite{altieri2015changepoint} proposed a method to find multiple unknown changepoints over time in
the inhomogeneous intensity of a spatio-temporal point process, allowing for spatial and temporal
dependence within segments but without specifically addressing the search for changepoints in the spatial covariates.

Our methodology consists of a two-step procedure: tessellation identification and tessellated model fitting. The first step involves a segmentation algorithm applied to geographically weighted estimates from an assumed local model. We employ the Simple Linear Iterative Clustering (SLIC) algorithm \citep{Achanta2012}, followed by a cluster analysis to merge superpixels with similar values, preventing over-segmentation. Superpixels are an effective tool for spatial segmentation, grouping cells with similar values into regions that maintain boundaries and structures while providing a desired level of homogeneity. 
Among the various methods available, SLIC stands out as one of the most prominent. In the second step, the resulting tessellation is treated as a categorical spatial variable in a log-linear Poisson point process model, encoded via dummy variables that indicate the tile to which each point belongs. This enables us to obtain region-wise covariate effects while maintaining compatibility with standard Poisson log-linear models.

An important advantage of our methodology is that it naturally addresses some common challenges in segmented regression models: the selection of the number and position of changepoints, here represented by the tiles of the tessellation. Indeed, the first step of the segmentation and clustering procedure automatically determines their number and position before the second step of model fitting. Although the proposed method does not explicitly address the problem of testing for the existence of a threshold, it does not require any prior knowledge or assumption regarding the number of abrupt spatial changes. In classical changepoint analysis, most fitting procedures focus on estimating the changepoint location and the associated covariate effects, typically under the assumption that a changepoint exists. This issue has been extensively studied \citep{feder1975log,beckman1979testing,ulm1991statistical}, highlighting that the distribution of the test statistic is often complex and depends on the alternative hypotheses, even in a nonspatial setting \citep{muggeo:03}. In contrast, our method relies on a clustering approach to determine the number of tiles in the tessellation by means of the higher Silhouette value. Consequently, although the results do not provide an explicit measure of uncertainty, the procedure will prefer a model with no changes in the parameters whenever the algorithm selects a single cluster of parameters' values.

We consider different formalisations of the proposed approach. The most general tessellated model assumes that each covariate is allowed to vary independently of the others across space, leading to potentially different tessellations for different predictors. Additionally, our proposed procedure, applied to the maps of the solely space-varying intercept (i.e., \lq \lq constant" intensity) faces the problem of feature detection. Indeed, one of the main interests of spatial point pattern analysis is identifying features surrounded by clutter. The conventional terminology is that a feature is a point of the pattern or process of interest, and clutter
(also called noise) consists of extraneous points that are not proper to the pattern of
interest. For example, the detection of surface minefields in an image captured by
a reconnaissance aircraft involves processing to generate a list of objects, among
which some could be mines while others may belong to different object categories \citep{allard1997nonparametric}. Similarly to \cite{Byers1998}, we are able to estimate this separation without making any assumptions about
the shape or number of features.
In other words, our approach represents a simple and intuitive method for estimating regions of different intensities in a point process.
Furthermore, we introduce a simpler version of the tessellated model, where a single tessellation is shared among all parameters, ensuring that the entire set of covariate effects changes coherently across the identified regions. 
The motivation from this further simplification 
stems from the need for simplification in the interpretation of the fitted model.

The proposed framework supports multiple applications,
and it can be conceptually extendable to geostatistics where the aim is not to model the point pattern intensity but a generic spatial response variable.  Furthermore, our methodological proposal lies the basis for many future extensions, including multitype Poisson point process models, more complex models based on the Poisson one, like the log-Gaussian Cox process models, and all of their spatio-temporal extensions. 

Through simulation studies, we assess the performance of our method, highlighting its dependence on the number of observed points and the magnitude of parameter changes. Applications to two examples where a model with region-wise parameters could be appropriate 
illustrate the proposal's potential for real-world spatial analysis. Furthermore, the analysis of the environmental problem of earthquake occurrence is addressed by taking into account the data of \cite{dangelo2021locall}.

The structure of the paper is as follows. After the preliminaries in Section \ref{sec:pre}, Section \ref{sec:examples} and Section \ref{sec:motiv} present two examples and the earthquake data that motivated this work. Then, 
Section \ref{sec:tess} introduces the proposed methodology, and  Section \ref{sec:sims} explores various simulation scenarios to understand the performance of the method in terms of tessellation identification, parameter estimation, and goodness-of-fit. Section \ref{sec:appl} provides the applications to the data introduced in Section \ref{sec:examples}. Moreover, the Greek seismic data are analysed through a tesselleted model in Section \ref{sec:greece}. Finally, new avenues for further methodological developments are addressed in Section \ref{sec:concl}.

\section{Preliminaries}\label{sec:pre}

\subsection{Spatial point processes}

We consider a (simple) point process $X=\{x_i\}_{i=1}^N$ 
(\cite{daley:vere-jones:08}), with points $x_i$ in the $2$-dimensional Euclidean space domain $W \in \mathbb{R}^2$. Formally, $X$ is a random element in the measurable space of locally finite point configurations/patterns $\mathbf{x}=\{(x_1,\ldots, x_n)\}$, $n\geq0$. A point location in the two-dimensional plane is denoted by a lowercase letter like $u$. Any location $u$ can be specified by its Cartesian coordinates $u = (u_1 ,u_2 )$ in such a way that we do not need to mention the
coordinates explicitly.
The first-order property of $X$ is described by the intensity function, defined as
\begin{equation*}
	\label{eq:spatial_th_int}
	\lambda(u) = \lim_{|\text{d}u| \rightarrow 0} \frac{\mathbb{E}[N(\text{d}u)]}{|\text{d}u|}
\end{equation*}
where $\text{d}u$ is an infinitesimal region that contains the point $u\in W$, $|\text{d}u|$ is its area and
$\mathbb{E}[N(\text{d}u)]$ denotes the expected number of events in $\text{d}u$.
When the intensity is constant, the process is called homogeneous. 
In the inhomogeneous case, the intensity is not constant in the study area but may depend, for instance, on the coordinates of points. 
A point process model, assuming independence,
is completely described by its intensity function $\lambda(u)$ \citep{daley:vere-jones:08}.

A general spatial log-linear Poisson model generalises both homogeneous and inhomogeneous models can be expressed as
\begin{equation}
	\label{eq:loglin}	\lambda(u)=\lambda(u,\boldsymbol{\theta}) = \exp\Biggl\{ \theta_0 
+ \sum_{j=1}^{p}\theta_jZ_j(u)\Biggr\},
\end{equation}
where $u \in W $, $\textbf{Z}(u)=\{Z_1(u), \ldots, Z_p(u)\}$ are  $p$ spatial covariates and $\boldsymbol{\theta}=(\theta_0,\theta_1, \ldots, \theta_p)$ are unknown  parameters. 
The estimation of the point process  parameters  is carried out through the maximization of the log-likelihood, defined as 
\begin{equation}
 \log \textrm{L}(\boldsymbol{\theta}) = \sum_{\textbf{x}_i \in \textbf{x}} \text{log} \lambda(\textbf{x}_i; \boldsymbol{\theta}) -\int_W\lambda(u; \boldsymbol{\theta})du.   
 \label{eq:loglik}
\end{equation}

\subsection{Local Poisson point process models}

When spatial log-linear models are used, the model parameters are usually assumed to be constant across the entire study region. This assumption may be too simplistic in contexts characterised by multi-scale and fractal features, like the seismic one, where the relationship between the intensity of earthquakes and other possible characteristics of the area where events occur vary spatially. In this context, the features can be described by a \textit{local} Poisson process with intensity
\begin{equation}
	\lambda(u)=\lambda\{u;\boldsymbol{\theta}(u)\}=\exp\Biggl\{ \theta_0(u) 
+ \sum_{j=1}^{p}\theta_j(u)Z_j(u)\Biggr\}
	\label{eq:eq281}
\end{equation}
with local coefficients being functions of $u$.
This is the local version of the global log-linear intensity \eqref{eq:loglin}.
For estimating local models as in Equation (\ref{eq:eq281}),
the likelihood at $u$ is used, being a weighted version of the original likelihood, with the greatest weight
on locations close to $u$.
Therefore, in the local context, the local log-likelihood associated with location $s$ (\cite{loader1999bandwidth}) is 
\begin{equation}
   \log \textrm{L}(s;\boldsymbol{\theta})=\sum_{i=1}^n w_h(\textbf{x}_i-s)\log\lambda(\textbf{x}_i;\boldsymbol{\theta})-\int_{W}\lambda(u;\boldsymbol{\theta})w_h(u-s)du,
   \label{eq:logL}
\end{equation}
where $w_h(u) = h^{-d}w(u/h)$ is a weight nonparametric function, and $h > 0$ is a smoothing bandwidth. It is not necessary to assume that $w_h$ is a probability density. Throughout the paper, we consider a kernel of fixed bandwidth  $h$.
In the local likelihood in Equation \eqref{eq:logL}, this is usually chosen by the cross-validation criterion   \citep{loader1999bandwidth}. The optimal bandwidth $h_{opt}$ maximises
$
	\textrm{LCV}(h) = \sum_i \log \lambda(\textbf{x}_i;\hat{\boldsymbol{\theta}}_i(\textbf{x}_i))-\int_D \lambda(u;\hat{\boldsymbol{\theta}}(u))du
$
where $\hat{\boldsymbol{\theta}}(u) = \hat{\boldsymbol{\theta}}(u,h)$ is the local estimate of $\boldsymbol{\theta}$ at location $u$ using bandwidth $h > 0$, and $\hat{\boldsymbol{\theta}}_i(\textbf{x}_i) = \hat{\boldsymbol{\theta}_i}(\textbf{x}_{i};h)$ is the corresponding `leave-one-out' estimate at the location $\textbf{x}_i$ computed from the data omitting $\textbf{x}_i$.

Maximizing the local likelihood for each fixed $u$ gives local parameter estimates $\hat{\boldsymbol{\theta}}(u)$, also known as \textit{geographically weighted estimates}. These fitted coefficients can be plotted as a function of spatial location $u$.  However, geographically weighted regression presents a major limitation. Its continuous nature makes it difficult to identify abrupt changes in the effect of covariates, as it imposes smooth transitions across space. 

The alternative we propose in this paper and introduce in the Section \ref{sec:tess} is \textit{tessellated regression}, which partitions the study region into distinct tiles and estimates regression coefficients separately for each of them. In other words, we assume that the covariates influencing the intensity of the observed point pattern have region-wise effects. This method aggregates data within tiles rather than estimating parameters continuously across space. Unlike geographically weighted regression, this approach allows for discrete changes in regression coefficients, making it possible to capture abrupt spatial variations in the effect of covariates. 
Therefore, while geographically weighted regression is useful for modeling smoothly varying spatial effects, it may fail when abrupt changes occur. In particular, \textit{tessellated regression} improves interpretability in spatial point process modeling. 

The key assumption in geographically weighted regression is that, for each spatial location $u \in W$, there is an unobserved spatial tessellation $W(u)$ containing $u$ where the global model is \textit{exactly} true with parameters $\boldsymbol{\theta}(u)$. If the support of the kernel centred at $u$ falls entirely inside $W(u)$, then the statistical properties of $\hat{\boldsymbol{\theta}}(u)$ are the same as they would be if the entire process followed the global model with constant parameter value $\boldsymbol{\theta} = \boldsymbol{\theta}(u)$ (\cite{baddeley:2017local}).
This motivates us to define the \textit{tessellated model} (Section \ref{sec:tess}) by means of the tessellations obtained from the maps of local estimates through a spatial segmentation algorithm.

\section{Examples}\label{sec:examples}

In this Section, we present two examples of point patterns for which a model with region-wise parameters could be more appropriate than both global and local Poisson models.

\subsection{Gordon Square data}

\begin{figure}[h]
    \centering
    \subfloat[Gordon Square data]{\includegraphics[width=.5\textwidth]{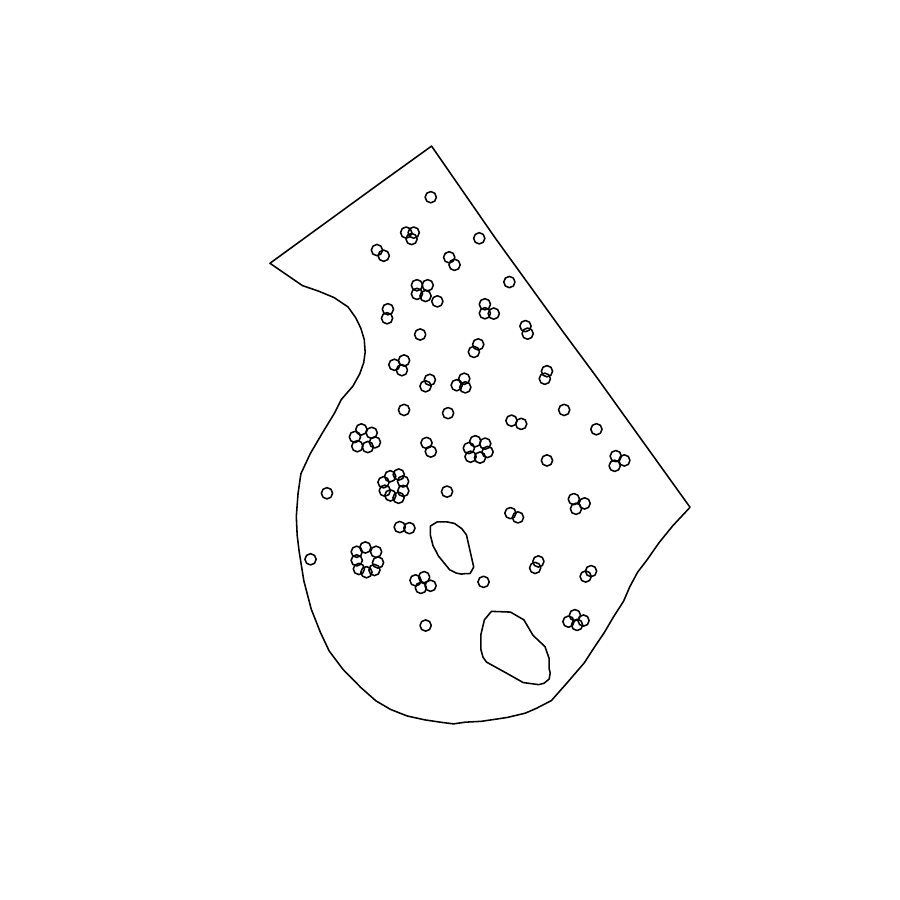}}
    \subfloat[Spatially varying intercept]{\includegraphics[width=.5\textwidth]{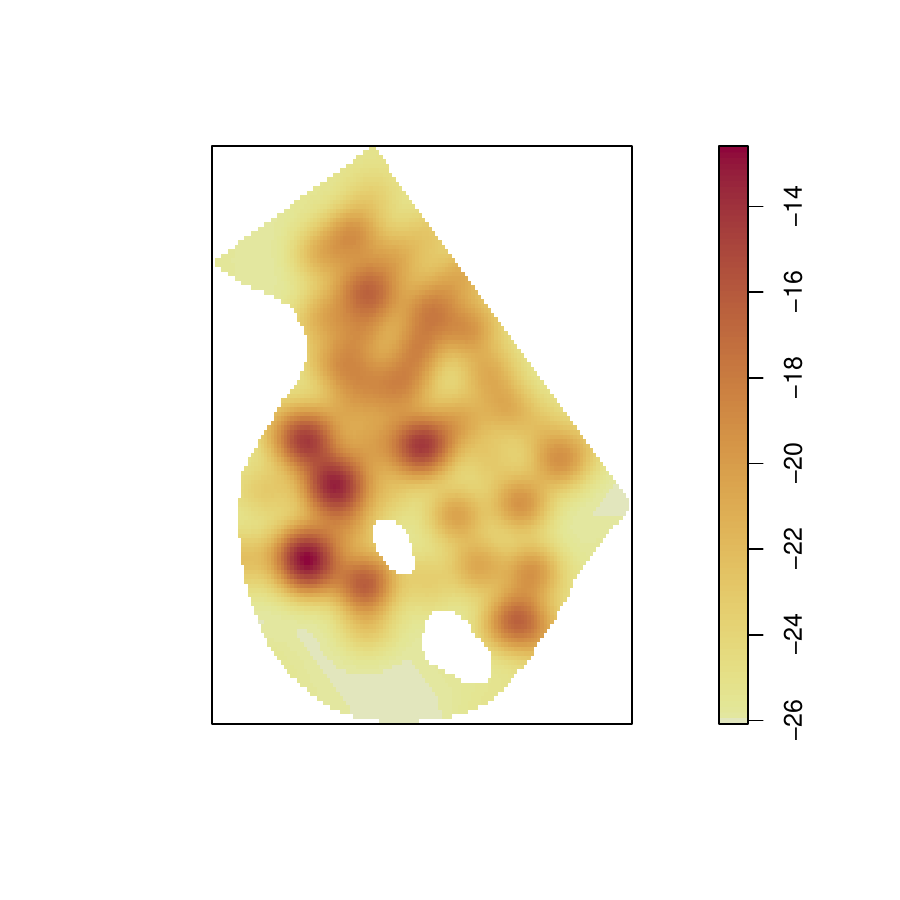}}
    \caption{(a) Gordon Square data. Locations of 99 people (circles) sitting on grass (gray shading)
in Gordon Square, London, UK on a sunny afternoon \citep{baddeley2013hybrids}. (b) Spatially varying intercept estimated from model   \ref{eq:gordon_local}.}
    \label{fig:gordon}
\end{figure}

The simplest case occurs when the aim is to detect changes in the space-varying homogeneous intensity of a local Poisson point process model.
For instance, Figure \ref{fig:gordon}(a) shows the spatial locations of people sitting on the grass in Gordon Square, London,
on a sunny afternoon \citep{baddeley2013hybrids}. The pattern appears to show spatial organization at several different
scales.
It would be possible to model the point pattern as a realisation of a
cluster or inhibitive process. However, the interest might lie in the will to estimate the locations of the clusters and treat them as \textit{feature} patterns.
With this aim in mind, the proposed approach could try segmenting the spatially varying \lq \lq homogeneous" intensity estimated by the following local model
\begin{equation}
    \label{eq:gordon_local}
\hat{\lambda}_{gordon}(u) = \exp\{ \hat{\theta}_0(u)\},
\end{equation}
to obtain the two constant intensities in the two subregions clearly defined by the higher intensity, that is, where at least two points form a spatial cluster (see Figure \ref{fig:gordon}(b)).

\subsection{Bronze filter data}

\begin{figure}[h]
    \centering
    \includegraphics[width=.75\textwidth]{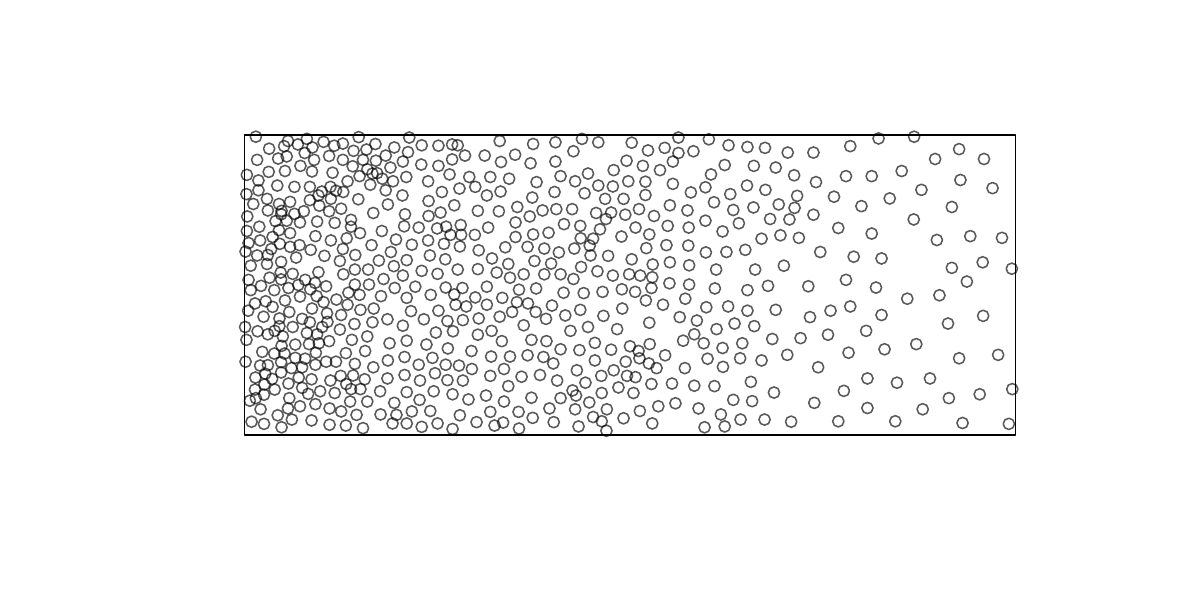}
    \caption{Spatially inhomogeneous pattern of circular section profiles of particles, observed in a longitudinal plane
section through a gradient sinter filter made from bronze powder \citep{bernhardt1997fundamental}.}
    \label{fig:bronze}
\end{figure}

To illustrate a case where the aim is to detected changes in the space-varying parameters of a covariate of a fitted local Poisson point process, Figure \ref{fig:bronze} shows centres of grains in a bronze filter \citep{bernhardt1997fundamental}.  
The pattern is clearly inhomogeneous with a negative effect of the x-coordinate. This effect however, is not constant, nor smoothly changing along the coordinate. The estimation of region-wise effects of the x-coordinate could help in finding the specific places where the effect of the coordinate changes.

\begin{figure}[h]
    \centering
    \includegraphics[width=\textwidth]{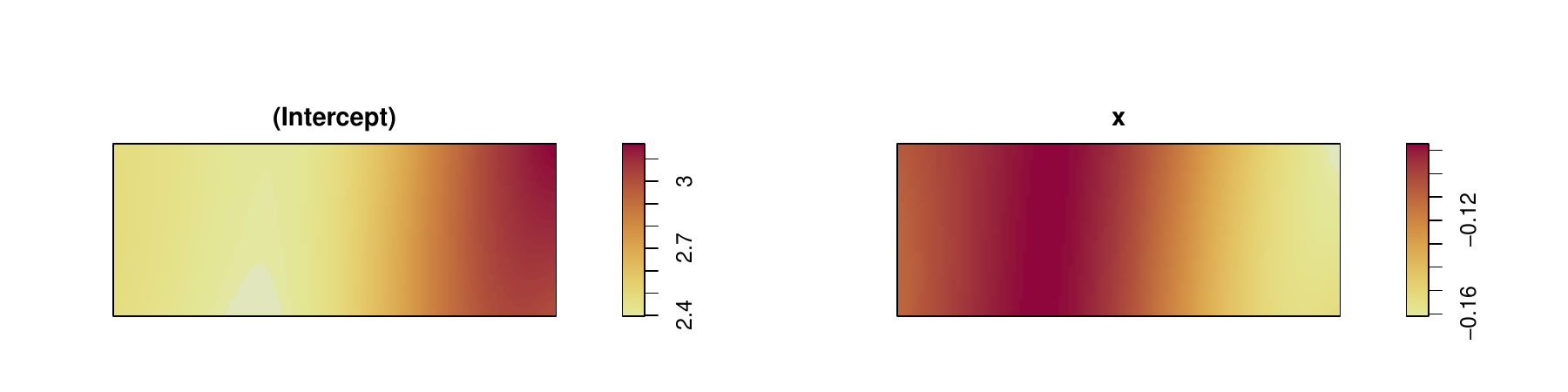}
    \caption{Spatially varying coefficients of model \eqref{eq:local_bronze} fitted to the Bronze filter data.}
    \label{fig:bronze_local}
\end{figure}

To have a first idea on the possible changes in the coordinate effect, Figure  \ref{fig:bronze_local} depicts the local coefficients estimated from the model
\begin{equation}
\hat{\lambda}_{bronze}(u) = \exp\{ \hat{\theta}_0(u) + \hat{\theta}_1(u)Long(u) \},
\label{eq:local_bronze}
\end{equation}
clearly displaying a sharp change in both the space-varying intercept and slope coefficient of the x-coordinate ($Long(u)$).  Ideally our methods should detect
the different types of patterns and estimate the boundary between them using only the point
location data. 

\section{Motivating problem: Greek seismicity data}\label{sec:motiv}

\begin{figure}[h]
	\centering
{\includegraphics[width=0.33\textwidth]{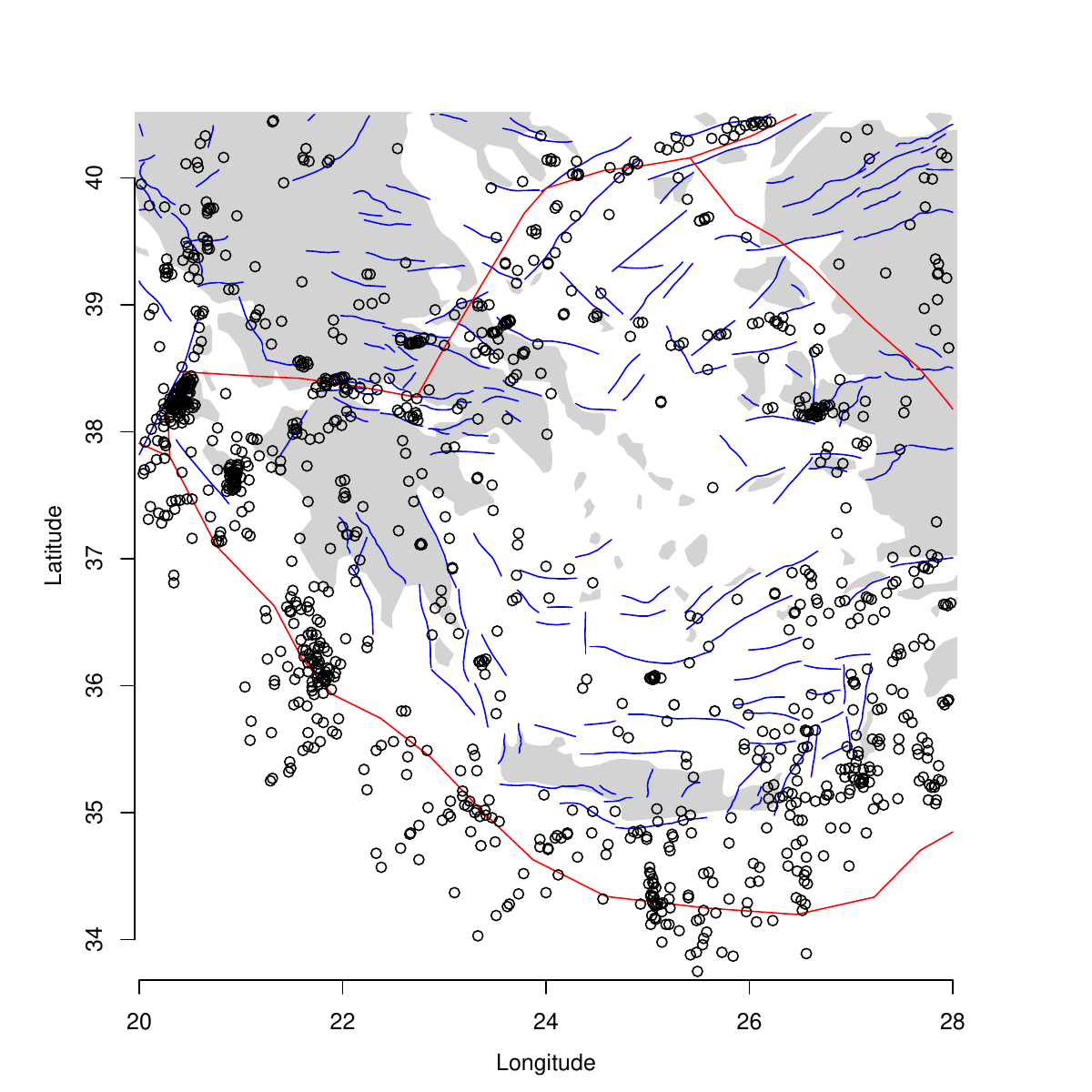}}
	\caption{Seismic point pattern.} 
	\label{fig:5}
\end{figure}

As anticipated, the motivation of this work arose from the will to describe the abrupt changes in the effects of seismic sources onto the earthquake phenomenon. In particular, we consider the same data analysed in \cite{siino2017spatial,dangelo2021locall,d2022locally,d2024minimum}, related to 1111 earthquakes that occurred in Greece between 2005 and 2014 with a magnitude greater than 4, coming from the Hellenic Unified Seismic Network (H.U.S.N.). These are in Figure 	\ref{fig:5}, with faults in blue and plate boundaries in red.

\begin{figure}[h]
	\centering
    	\subfloat[$\hat{\theta}_0(u)$]{\includegraphics[width=0.33\textwidth]{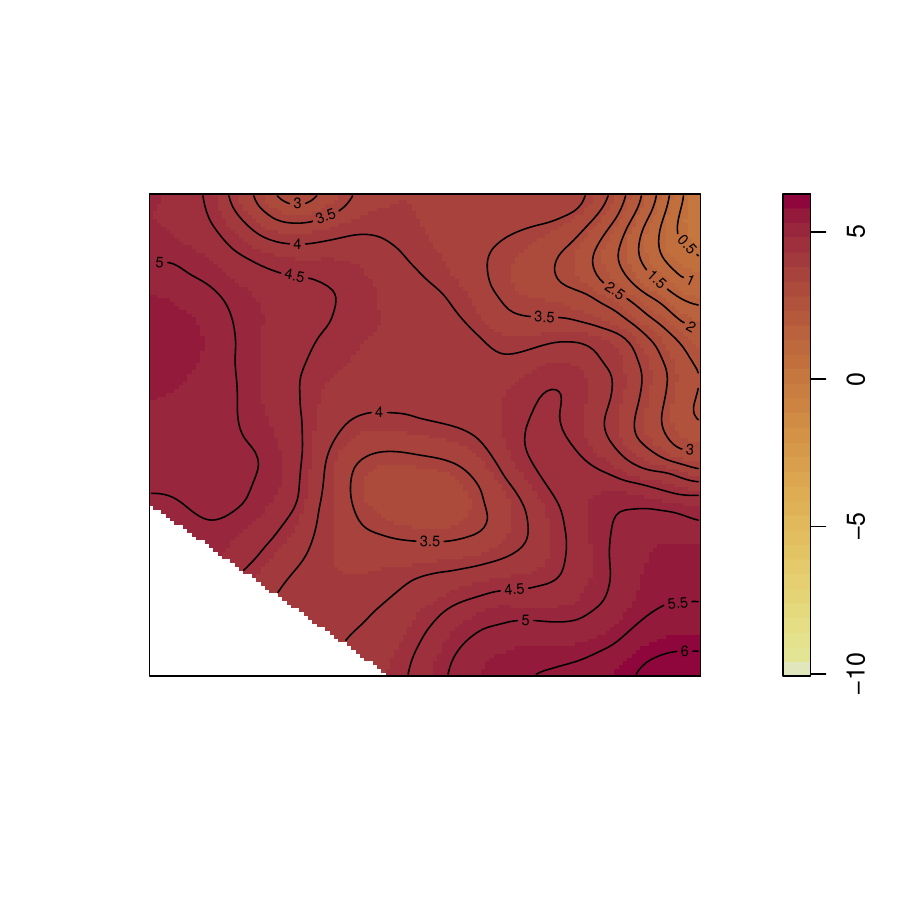}}
	\subfloat[$\hat{\theta}_1(u)$]{\includegraphics[width=0.33\textwidth]{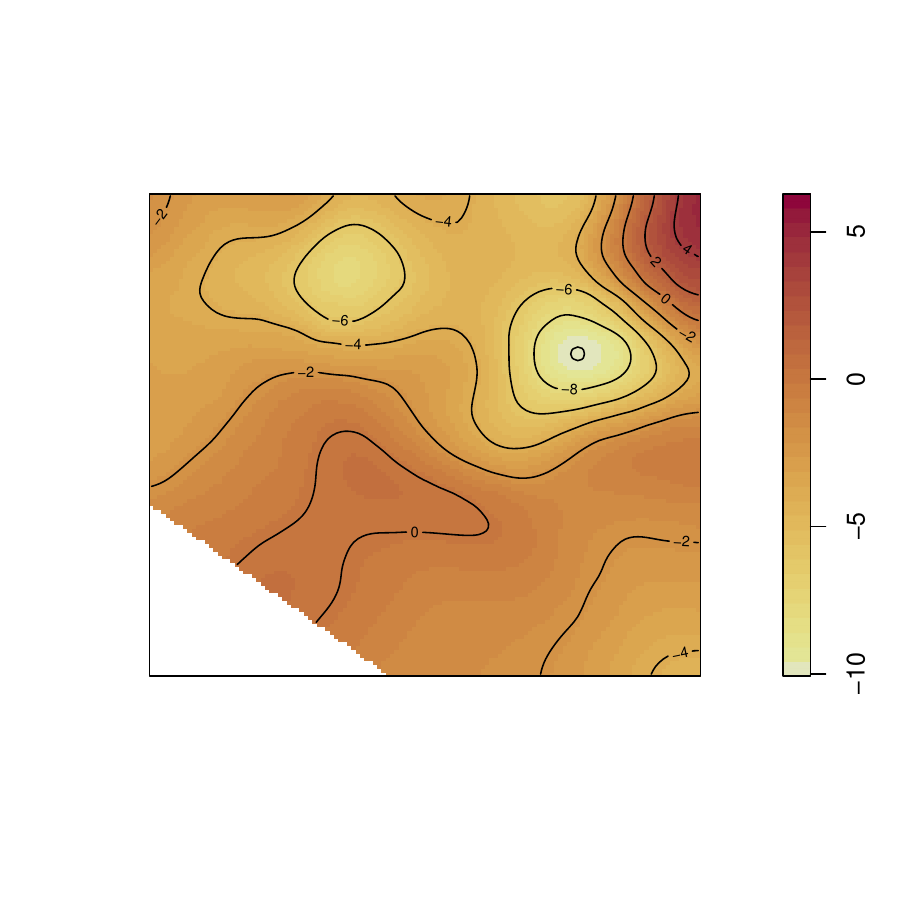}}
    	\subfloat[$\hat{\theta}_2(u)$]{\includegraphics[width=0.33\textwidth]{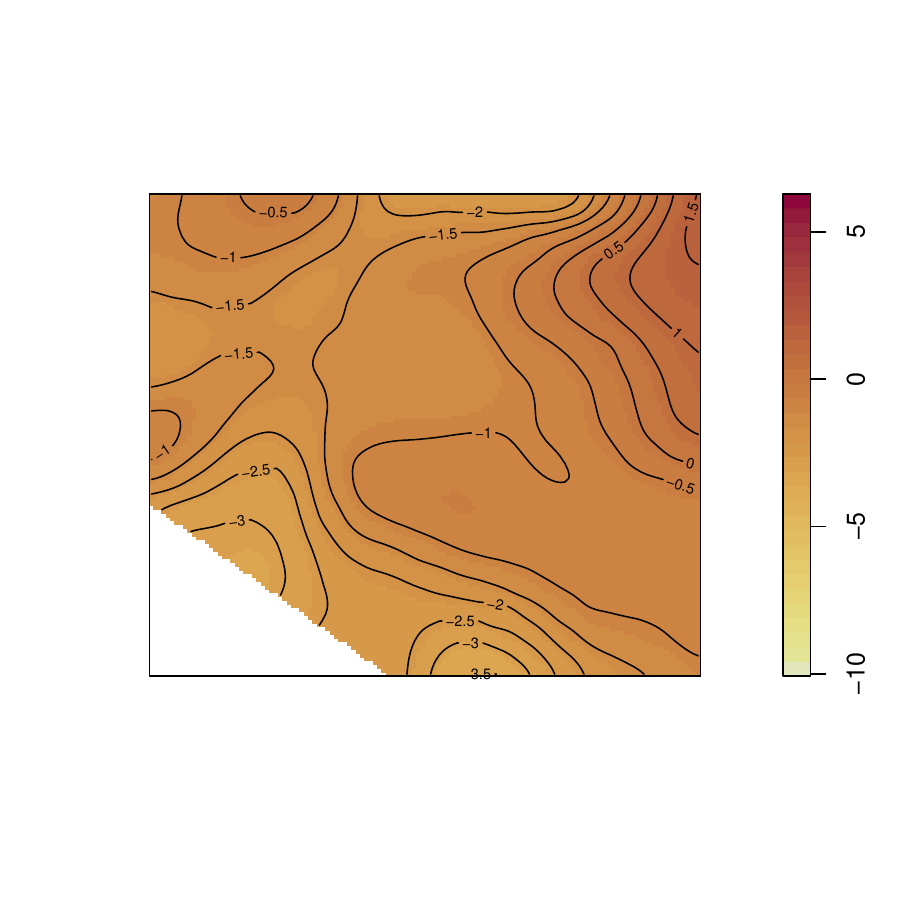}}
        \caption{Spatial varying coefficients $\hat{\theta}_1(u)$, $\hat{\theta}_2(u)$ and $\hat{\theta}_3(u)$ of the local Poisson Model in Equation \eqref{eq:add}, i.e., the coefficients of the intercept and the variables $D_{f}(u)$ and $D_{pb}(u)$, respectively.}
	\label{fig:gg10}
\end{figure}

The aim is to understand whether the two spatial variables \textit{Distance from the plate boundary} ($D_p(u)$) and \textit{Distance from the faults} ($D_f(u)$)
computed as the Euclidean distances from the spatial location $u$ of events and the map of geological information, affect the intensity of the seismic process. This application is particularly interesting as it also deals with changes in the space-varying parameters of an external covariate in local Poisson point processes. Note that \cite{dangelo2021locall} already proved the variables' effects to be local (see Figure \ref{fig:gg10}) through the model
\begin{equation}
   \hat{\lambda}_{greece}(u) = \exp\{ \hat{\theta}_0(u) + \hat{\theta}_1(u)D_p(u) + \hat{\theta}_2(u)D_f(u)\}.
   \label{eq:add}
\end{equation} For this reason, we are interested in assessing whether a model with space-varying parameters could be appropriate while also clearly defining the regions where these effects change.

\section{Tessellated spatial Poisson point process models}\label{sec:tess}

The \textit{tessellated} model we propose is, therefore, expressed as 
\begin{equation}
	\label{eq:loglin_segmented}	\lambda(u)=\lambda(u,\boldsymbol{\beta},\boldsymbol{\gamma}) = \exp\Biggl\{ \beta_0 + \sum_{j=1}^{p}\beta_jZ_j(u) +\sum_{k=1}^{q-1}\gamma_{0k}W_k(u)+\sum_{j=1}^{p}\sum_{k=1}^{q-1}\gamma_{jk}Z_j(u)W_{Z_j,k}(u) \Biggr\},
\end{equation}
where the $\boldsymbol{\beta}$ parameters are the effects of the non-tessellated spatial covariates and the $\boldsymbol{\gamma}$ parameters are those referred to the change in the effects of the covariates in the tiles identified by the tessellation $W(u)$.
A tessellation refers to the division of space into non-overlapping regions or cells, often used to analyse the spatial structure of point patterns. 
The tessellation $W(u)$ of the variable $Z_j(u)$ is indicated by $W_{Z_j}(u)$, and it has $q$ tiles, encoded in the model using  $q-1$ dummy variables. They can be interpreted as categorical variables indicating which tile of the tessellation $u$ belongs to.
Inference on $\boldsymbol{\gamma}_j$, and on the tessellation of the $j$th covariate in turn, can be carried out, for instance, through the Likelihood Ratio Test (LRT), since model \eqref{eq:loglin} and \eqref{eq:loglin_segmented} are nested. We decided to denote the non-tessellated effects in model \eqref{eq:loglin_segmented} by $\boldsymbol{\beta}$ to avoid confusion with $\boldsymbol{\theta}$ of model \eqref{eq:loglin} in the applications.

A special case of \eqref{eq:loglin_segmented}, 
is defined as follows
\begin{equation}
	\label{eq:loglin_embedded}
\lambda(u)=\lambda(u,\boldsymbol{\beta},\boldsymbol{\gamma}) = \exp\Biggl\{ \beta_0 + \sum_{j=1}^{p}\beta_jZ_j(u) + \sum_{k=1}^{q-1}\gamma_{0k}W_k(u)+ \sum_{j=1}^{p}\sum_{k=1}^{q-1}\gamma_{jk}Z_j(u)W_k(u) \Biggr\}.
\end{equation}
Basically, this represents the case of the tessellated model \eqref{eq:loglin_segmented} with $W(u)$ common to every variable, ensuring that the entire set of covariate effects changes coherently across the identified regions. This model is even easier to interpret with respect to the tessellated one, and therefore, if the data are better described by this model, one further gains in interpretability of the phenomenon under analysis. 

We employ the \texttt{supercells} (\cite{nowosad2022extended}) R (\cite{R}) package for obtaining the tessellation and \texttt{spatstat} (\cite{spat}) for the model fitting.

Of course, the first step and main challenge here is the identification of the tessellations $W(u)$, discussed below. 

\subsection{Tessellation identification}


Superpixels are an effective tool for spatial segmentation, grouping cells with similar values into regions that maintain boundaries and structures while providing a desired level of homogeneity. These groupings not only enrich the information content compared to single cells but also accelerate subsequent processing. Among the various methods available, the  Simple Linear Iterative Clustering (SLIC) algorithm \citep{Achanta2012} stands out as one of the most prominent. 

SLIC begins by initializing cluster centers at regular intervals and assigning each cell to the nearest cluster center based on a combined distance metric that incorporates both spectral and spatial proximity. 
The distance $D$ between a cell and a cluster center is defined as
\begin{equation}
    \label{eq:dist}
    D = \sqrt{d_c^2 + g \frac{d_s^2}{S^2}},
\end{equation}
where $d_c$ is the color (spectral) distance, $d_s$ is the spatial (Euclidean) distance, $S$ is the interval between initial cluster centers, and $g$ is the compactness parameter that balances the influence of spectral and spatial distances.  Higher values of $g$ produce regularly shaped superpixels, whereas lower values result in more spatially adapted, irregular shapes.

The color distance is calculated as
$$
d_c = \sqrt{\sum_{p \in B} \left( I(x_i, y_i, sp) - I(x_j, y_j, sp) \right)^2},
$$
where $B$ is the set of spectral bands, and $I(x_i, y_i, sp)$ represents the intensity value of the spectral band $sp$ at location $(x_i, y_i)$. The spatial distance is given by
$$d_s = \sqrt{(x_j - x_i)^2 + (y_j - y_i)^2}.$$

The algorithm iteratively updates cluster centers and assignments until convergence, typically achieved after 4–10 iterations.

The SLIC algorithm provides an efficient and effective method for generating superpixels, but it may result in over-segmentation when the number of superpixels is too high. Indeed, if the number of tiles is too small, the model may not capture all the changes in the relationship between the variables, resulting in bias and reduced model fit. On the other hand, if the number of tiles is too large, the model may be overfitting the data and may not generalise well to new data. 
We, therefore, propose to carry out hierarchical clustering for the $j$th covariate as follows.
\begin{enumerate}
    \item Compute a distance matrix using the distance metric  \eqref{eq:dist} on the maps of the estimated local parameters $\hat{\boldsymbol{\theta}}_j(u)$ coming from the local model 	\eqref{eq:eq281}.
    \item Perform hierarchical clustering using a method like Ward’s linkage on the resulting superpixels' values.
    \item Cut the dendrogram to define the desired number of clusters according to the highest silhouette value.
    \item Assign cluster labels to the superpixels to form the tiles to obtain the tessellation $W_{Z_j}(u)$.
\end{enumerate}
For the simplest case of model \eqref{eq:loglin_embedded}, the SLIC algorithm is applied to the multiple rasters of the maps of all the parameters in the corresponding local model. On the other hand, the main perk of model \eqref{eq:loglin_segmented}	over model \eqref{eq:loglin_embedded} is that the interval between initial cluster centres $S$ and the compactness parameter $g$ can be tuned separately for each covariate.

\subsection{Model fitting}

Once the tessellation for each covariate is obtained, the \textit{tessellated} model 
\eqref{eq:loglin_segmented}	
can be fitted.

Any tessellated model can be fitted through the established Berman-Turner technique \citep{berman1992approximating}, also called \textit{quadrature scheme}. 
Renaming the data points as $x_1,\dots , x_n$ with $({u}_i,t_i) = x_i$ for $i = 1, \dots , n$, then generate $m$  additional ‘‘dummy points’’ $({u}_{n+1},t_{n+1}) \dots , ({u}_{m+n},t_{m+n})$ to
form a set of $n + m$ cubature points (where $m > n$). Then we determine cubature weights $a_1, \dots , a_m$
so that the integral in \eqref{eq:loglik} can be approximated by a Riemann sum
$    \int_W \lambda({u};\boldsymbol{\theta})\text{d}u \approx \sum_{k = 1}^{n + m}a_k\lambda({u}_{k};\boldsymbol{\theta})$
where $a_k$ are the cubature weights such that $\sum_{k = 1}^{n + m}a_k = l(W)$ where $l$ is the Lebesgue measure.

The log-likelihood \eqref{eq:loglik} of the template model can be approximated by
\begin{equation}
\begin{split}
        \log L(\boldsymbol{\theta})   \approx &
\sum_k
\log \lambda({u}_k; \boldsymbol{\theta}) +
\sum_k
(1 - \lambda({u}_k; \boldsymbol{\theta}))a_k
 = 
\sum_k
e_k \log \lambda({u}_k; \boldsymbol{\theta}) + (1 - \lambda({u}_k; \boldsymbol{\theta}))a_k
\end{split}
\label{eq:approx0}
\end{equation}

where $e_k$ is the indicator that equals $1$ if $(u_k,t_k)$ is a data point.
Writing $y_k = e_k/a_k$ \eqref{eq:approx0} becomes $
    \log L(\boldsymbol{\theta}) \approx
\sum_k
a_k
(y_k \log \lambda({u}_k, t_k; \boldsymbol{\theta}) - \lambda({u}_k, t_k; \boldsymbol{\theta}))
+
\sum_k
a_k.$

Apart from the constant $\sum_k a_k$, this expression is formally equivalent to the weighted log-likelihood of
a Poisson regression model with responses $y_k$ and and weights $a_k$. This can be
maximised using standard GLM software.
Another option not detailed in this work is the logistic spatial regression \citep{baddeley2014logistic}.

\section{Simulation study}\label{sec:sims}
This section is devoted to some numerical studies to understand the performance of the method in terms of tessellation identification,  parameter estimation, and goodness-of-fit. 

\subsection{Tessellation identification}

We consider point processes with region-wise constant intensity as follows
\begin{equation}
\lambda = \exp\{\beta_0  +
		\gamma_0 W(u)\}		
	\label{eq:syst1}
\end{equation}
with \begin{equation*}
W(u) = W(x,y)= 
	\begin{cases}
		1& \text{if } x > y \\
		0  & \text{otherwise} 
	\end{cases}
\end{equation*}
meaning that we assume a constant intensity and only two tiles where this changes.
To understand the ability of our proposal to correctly identify $W(u)$, we simulate 100 patterns from intensity \eqref{eq:syst1} and compute the percentage of times the algorithm correctly identifies the two tiles.

Table \ref{tab:dati} presents the results for different values of expected number of points $\mathbb{E}[N]$, and parameters \( \beta_0 \), and \( \gamma_0 \). 

\begin{table}[h]
    \caption{Percentage of times the algorithm correctly identifies the two tiles in \eqref{eq:syst1}, over 100 simulations.}
    \centering
    \begin{tabular}{r|rr|r}
        \toprule
        $\mathbb{E}[N]$ & $\beta_0$ & $\gamma_0$ & percentage  \\
        \midrule
   \multirow{2}{*}{500}  &4.61  & 5.99  &  0.78 \\
          &3.91   & 6.11  &  0.79 \\
        \midrule
           \multirow{3}{*}{600}   &5.3  & 5.99  &  0.74 \\
          &4.61  & 6.21   & 0.86 \\
          &1.61    & 6.31    & 0.91 \\
        \bottomrule
    \end{tabular}
    \label{tab:dati}
\end{table}

When \( \mathbb{E}[N] = 500 \), the identification percentage remains relatively stable, while for \( \mathbb{E}[N] = 600 \), the accuracy shows greater variation, suggesting that increasing the expected number of points does not necessarily lead to an improvement in the tessellation identification performance. Instead, the accuracy appears to be more sensitive to the values of \( \beta_0 \) and \( \gamma_0 \). 
In particular, examining the ratio \( \beta_0/ \gamma_0 \), we observe that when \( \beta_0 \) and \( \gamma_0 \) are relatively close, the identification percentage is lower. Conversely, when the gap increases, the identification accuracy reaches its highest value. This suggests that a larger separation between \( \beta_0 \) and \( \gamma_0 \) may contribute to better tessellation identification performance.

\subsection{Parameter estimation}

Assuming now the tessellation $W(u)$ to be known, we study the performance of the algorithm in terms of point estimation. To this aim, we explore both the cases of a region-wise constant intensity and the case of an inhomogeneous intensity depending on a spatial region-wise covariate. We illustrate the performance of the method by simulating patterns from such intensities and by fitting our proposed tessellated model. We report the mean and standard deviations of the estimated parameters, together with their Mean Squared Error (MSE), averaged over 100 simulations.

In the first scenario, the intensity is the same as in equation \eqref{eq:syst1}, but with varying expected number of points and ratios between $\beta_0$ and $\gamma_0$.  Table \ref{tab:sims2} presents the results. 

\begin{table}[h]
\centering
\caption{\label{tab:sims2} Mean (sd) and MSE of the parameters, averaged over the 100 simulated point patterns generated assuming intensity \eqref{eq:syst1}.}
\begin{tabular}{c|cc|cc|cc}
  \toprule
    & \multicolumn{2}{c|}{Parameters}  & \multicolumn{2}{c|}{$\hat{\beta}_0$}   & \multicolumn{2}{c}{$\hat{\gamma}_0$}   \\
      \midrule
   $\mathbb{E}[N]$ & $\beta_0$ & $\gamma_0$  & mean (sd)  & MSE & mean (sd) & MSE  \\ 
  \midrule
   \multirow{4}{*}{100}&3.91&3.91&3.85 (0.19)& 0.042& 3.93 (0.21)& 0.041 \\
  &3.69 &4.09 &3.6 (0.29)&0.090& 4.07 (0.21)&0.042\\  
   &3.00 & 4.38 & 2.89 (0.34)& 0.128 & 4.36 (0.16)&0.025 \\ 
 &2.30 &4.50&2.23 (0.41)&0.171& 4.47 (0.17)& 0.031 \\
   \midrule
   \multirow{4}{*}{500}&5.52&5.52&5.47  (0.09)&0.011& 5.52 (0.09)&0.008 \\
     &5.30 & 5.70 &5.26 (0.1)&0.012& 5.71 (0.08) &0.006 \\
      &4.61 &5.99& 4.58 (0.15)&0.022& 5.99 (0.08)&0.007 \\ 
 &3.91 &6.11 &3.94 (0.2)&0.040& 6.12 (0.07)&0.004 \\
   \midrule
   \multirow{4}{*}{1000}& 6.21& 6.21&6.18 (0.07)&0.006& 6.21 (0.07)&0.004 \\
 & 5.99& 6.40&5.95 (0.08)&0.008& 6.4 (0.05)&0.002 \\  
  &5.30&6.68& 5.28 (0.1)&0.011& 6.68 (0.05)&  0.003 \\ 
 &4.61 &6.80 &4.62 (0.14) & 0.020 & 6.81 (0.04) & 0.002\\
   \bottomrule
\end{tabular}
\end{table}

As \(\mathbb{E}[N]\) increase, the estimates of both parameters become more accurate, with means approaching the true values and the standard deviations and MSE decreasing. 
The ratio between the true values of \(\beta_0\) and \(\gamma_0\) also plays a role in the estimation accuracy. When the two parameters are closer in value, both parameters are estimated precisely with very low MSE. However, as the difference between \(\beta_0\) and \(\gamma_0\) increases, the estimate of \(\beta_0\) shows more variability, resulting in a higher MSE for \(\hat{\beta}_0\), while \(\hat{\gamma}_0\) remains more stable. Overall, the difference does not appear much relevant. We can, therefore, conclude that increasing \(\mathbb{E}[N]\) leads to a reduction in estimation error, and a more balanced ratio between the true parameters could yield more precise estimates with lower MSE, even though the improvement is quite small.

We now proceed with the second scenario, that is, the one where the intensity function has the following form
\begin{equation}
\lambda(u) = \exp\{\beta_0 + \beta_1Z(u)  + \gamma_1Z(u) W(u)\}
	\label{eq:syst2}
\end{equation}
with $Z(u)$ a spatial covariate generated from a Gaussian Radom Field with zero mean.
Equation \eqref{eq:syst2} implies that $Z(u)$ has region-wise constant effect, and, therefore, has an effect ($\beta_1$) on the intensity when $x < y$ and another ($\gamma_1$) otherwise. 
Table \ref{tab:sims3} displays the mean, standard deviations, and MSE of the estimated parameters \(\hat{\beta}_0\), \(\hat{\beta}_1\), and \(\hat{\gamma}_1\), averaged over 100 simulated point patterns generated according to the intensity function in equation \eqref{eq:syst2}.

\begin{table}[h]
\centering
\caption{\label{tab:sims3} Mean (sd) and MSE of the parameters, averaged over the 100 simulated point patterns generated assuming intensity \eqref{eq:syst2}.}
\begin{tabular}{c|ccc|cc|cc|cc}
  \toprule
    &\multicolumn{3}{c|}{Parameters} & \multicolumn{2}{c|}{$\hat{\beta}_0$}    & \multicolumn{2}{c|}{$\hat{\beta}_1$}    & \multicolumn{2}{c}{$\hat{\gamma}_1$}  \\
      \midrule
   $\mathbb{E}[N]$ & $\beta_0$ & $\beta_1$ & $\gamma_1$ & mean (sd)  & MSE & mean (sd) & MSE & mean (sd) & MSE  \\ 
  \midrule
100&3.15&1&2& 3.34 (0.16) &0.061&0.93 (0.17)& 0.033& 1.56 (0.07)&0.201\\
500&4.755&1&2&4.96 (0.09)&0.052& 0.87 (0.08)& 0.023& 1.58 (0.04)& 0.178\\
1000&5.5&1&2&5.71 (0.06)&0.045& 0.89 (0.05)&0.016 & 1.61 (0.03)&0.156\\  
   \midrule
100&1.15&1.5&3&1.95 (0.28)& 0.718& 1.29 (0.34)&0.155& 2.16 (0.10)&0.718\\  
500&2.75&1.5&3&3.15 (0.14)&  0.176& 1.52 (0.12)&0.015 & 2.45 (0.05)&0.304\\
1000&3.45&1.5&3&3.92 (0.09)&0.226& 1.46 (0.09)& 0.009& 2.43 (0.04)&0.329\\  
   \midrule
100&1.2&1&3&1.99 (0.27) &0.692& 1.10 (0.38)& 0.153& 2.15 (0.10)& 0.724  \\ 
500&2.775&1&3&3.17 (0.15)&0.176& 1.30 (0.18) &0.125&2.45 (0.06)&0.302\\
1000&3.5&1&3&3.93 (0.11)& 0.196& 1.27 (0.11)&0.083& 2.44 (0.04)&0.319\\ 
   \bottomrule
\end{tabular}
\end{table}

 As in the previous scenario, the results show that increasing \(\mathbb{E}[N]\) leads to more accurate estimates.
Also in this scenario, when the true values of \(\beta_1\) and \(\gamma_1\) are close, the estimates for \(\hat{\beta}_1\) are relatively precise even for smaller \(\mathbb{E}[N]\), and the MSE is lower compared to cases where \(\beta_1\) and \(\gamma_1\) differ more evidently. However, when the values of \(\beta_1\) and \(\gamma_1\) differ more, the estimates for \(\hat{\gamma}_1\) become less accurate, with higher MSE and greater variability in the estimates, especially for smaller values of \(\mathbb{E}[N]\). 
This suggests that a greater disparity between \(\beta_1\) and \(\gamma_1\) makes estimation more challenging, particularly in smaller samples. As \(\mathbb{E}[N]\) increases, the estimates become more stable, and the MSEs decrease, but the influence of the ratio between the parameters remains slightly visible in the results.

Finally, we also explore the case of all model parameters changing within the same tiles, 
assuming the following in intensity 
\begin{equation}
\lambda(u) = \exp\{\beta_0 + \beta_1Z(u) + \gamma_0 W(u) + \gamma_1 Z(u)W(u)\}
	\label{eq:syst3}
\end{equation}
basically signifying that all the parameters of an inhomogeneous model change according to the tiles of $W(u)$. Formally, we are considering an intensity of the form of equation 	\eqref{eq:loglin_embedded}. With respect to intensity \eqref{eq:syst2}, we are now also assuming different intercepts ($\beta_0$ and $\gamma_0$) in the two tiles.

Table \ref{tab:sims4} provides the mean, standard deviations, and MSE of the estimated parameters \(\hat{\beta}_0\), \(\hat{\beta}_1\), \(\hat{\gamma}_0\), and \(\hat{\gamma}_1\), averaged over 100 simulated point patterns generated according to the intensity function in equation \eqref{eq:syst3}.

\begin{table}[h]
\centering
\caption{\label{tab:sims4} Mean (sd) and MSE of the parameters, averaged over the 100 simulated point patterns generated assuming intensity \eqref{eq:syst3}.}
\resizebox{\textwidth}{!}{\begin{tabular}{c|cccc|cc|cc|cc|cc}
  \toprule
    & \multicolumn{4}{c|}{Parameters} &\multicolumn{2}{c|}{$\hat{\beta}_0$}   & \multicolumn{2}{c|}{$\hat{\beta}_1$}   & \multicolumn{2}{c|}{$\hat{\gamma}_0$}  & \multicolumn{2}{c}{$\hat{\gamma}_1$}   \\
      \midrule
   $\mathbb{E}[N]$ & $\beta_0$ & $\beta_1$ & $\gamma_0$  & $\gamma_1$ & mean (sd)  & MSE & mean (sd) & MSE & mean (sd) & MSE & mean (sd) & MSE  \\ 
  \midrule
  100 &2.5 & 0.50& 4.95& 0.75 & 2.51 (0.44) &0.192&  5.02 ( 0.12) &0.019& 0.50 (0.40) & 0.161& 0.61 (0.08)&
0.026\\
  500 & 3.40&0.50 &6.64 & 0.75 & 3.52 (0.26) &0.081& 6.69 (0.05) &0.005& 0.50 (0.24)  &0.058& 0.60 (0.03) & 0.023\\
  1000 & 3.74&0.50 & 7.31& 0.75 & 3.99 (0.20) & 0.100& 7.34 (0.03) &0.002& 0.51 (0.16) &0.025& 0.63 (0.02) & 0.014\\
   \bottomrule
\end{tabular}}
\end{table}

As in the previous scenarios, increasing the expected number of points \(\mathbb{E}[N]\) results in improved estimates. The MSE for both \(\hat{\beta}_0\) and \(\hat{\gamma}_0\) are relatively low even for \(\mathbb{E}[N] = 100\), with the mean values being very close to the true values. The estimates for \(\hat{\beta}_0\) and \(\hat{\gamma}_0\) improve further as \(\mathbb{E}[N]\) increases, although the MSE remain relatively stable across the larger sample sizes. This suggests that the parameters \(\beta_0\) and \(\gamma_0\) are relatively easy to estimate, even with smaller samples, when the values are reasonably close to each other.
The estimates for \(\hat{\beta}_1\) and \(\hat{\gamma}_1\) show a similar trend, with their MSE decreasing as \(\mathbb{E}[N]\) increases. 
The estimation of \(\hat{\beta}_1\) looks less sensitive to sample size compared to \(\hat{\gamma}_1\), but both parameters are more stable with larger sample sizes.

\subsubsection{Goodness-of-fit}

{This section is devoted to the study of the performance of the proposed method in terms of overall goodness-of-fit. We compare the proposed method to both the global and the local models in terms of Mean Integrated Squared Error (MISE).

Let $\lambda(\cdot)$ denote the intensity function of the generating process. 
Given an intensity estimator $\hat{\lambda}(\cdot) = \hat{\lambda}(\cdot;X)$, when the true intensity is known, the goodness-of-fit of $\hat{\lambda}(\cdot)$ may be measured by 
$$
\mathrm{MISE}_{\hat{\lambda}} = \mathbb{E}\Biggl[\int_W\big(\hat{\lambda}(u) - \lambda(u)\big)^2du\Biggl]
=
\int_W\mathrm{Var}\left(\hat{\lambda}(u)\right)du
+
\int_W\mathrm{Bias}(\hat{\lambda}(u))^2du
,
$$
where $\mathrm{Bias}(\hat{\lambda}(u)) = \mathbb{E}[\hat{\lambda}(u)] - \lambda(u)$. 
Since it is hard to compute $\mathrm{MISE}_{\hat{\lambda}}$ analytically for a given intensity estimator $\hat{\lambda}$, we estimate the MISE by $$\mathrm{MISE}_{\hat{\lambda}} = \frac{1}{n}\sum_{i=1}^n \int_W\big(\hat{\lambda}(u;\mathbf{x}_i) - \lambda(u)\big)^2du.$$

For each of the scenarios explored previously, we fit the tessellated model and compare its MISE with those of the corresponding global and local parameters. For the former, we fit the model with fixed effects coefficients and without any tessellation. For the latter, we fit the corresponding local model with space-varying parameters. The aim is to prove the best goodness-of-fit of the tessellated model when the change in the model parameters is actually abrupt.\\
Table \ref{tab:sims2mise} presents the MISE for three different models—global, local, and tessellated—averaged over 100 simulated point patterns generated with intensity function as defined in equation \eqref{eq:syst1}.

\begin{table}[h]
\centering
\caption{\label{tab:sims2mise} MISE for the global, local, and tessellated fitted models, averaged over the 100 simulated point patterns generated assuming intensity \eqref{eq:syst1}.}
\begin{tabular}{c|cc|rrr}
  \toprule
    & \multicolumn{2}{c|}{Parameters}& \multicolumn{3}{c}{MISE}     \\
      \midrule
   $\mathbb{E}[N]$ & $\beta_0$ & $\gamma_0$  & global  & local & tessellated  \\ 
  \midrule
   \multirow{4}{*}{100}&3.91&3.91&43&92&77\\
  &3.69 &4.09 &165&144&123\\  
   &3.00 & 4.38 &963&494&130 \\ 
 &2.30 &4.50&1692&847&185\\
   \midrule
   \multirow{4}{*}{500}&5.52&5.52&242&384&518\\
     &5.30 & 5.70 &2793&1543&600  \\
      &4.61 &5.99& 23250&10852&1525\\ 
 &3.91 &6.11 &18662&41041&2062\\
   \bottomrule
\end{tabular}
\end{table}

The table clearly shows the better performance of the tessellated models in all the scenarios. The only expectation is given when the ratio between $\beta_0$ and $\gamma_0$ is equal to 1, actually corresponding to a constant intensity. Indeed, in such cases, the lower MISE is correctly obtained by the global model. In all of the remaining scenarios,  however, the tessellated model proves its superiority compared to both the global and the local model. This result is even more evident as the number of expected number of points increases.

Tables \ref{tab:sims3mise} and \ref{tab:sims4mise} 
report the results for the other scenarios previously considered, always showing the best fitting performance achieved by the tessellated model, always followed by the global one and then by the local one.

\begin{table}[h]
\centering
\caption{\label{tab:sims3mise} MISE for the global, local, and tessellated fitted models, averaged over the 100 simulated point patterns generated assuming intensity \eqref{eq:syst2}.}
\begin{tabular}{c|ccc|rrr}
  \toprule
    & \multicolumn{3}{c|}{Parameters}& \multicolumn{3}{c}{MISE}\\
      \midrule
   $\mathbb{E}[N]$ & $\beta_0$ & $\beta_1$ & $\gamma_1$ & global & local& tessellated  \\ 
  \midrule
100&3.15&1&2&152361 &246243&116831\\
500&4.755&1&2&3787483&4783515&2674396\\
   \midrule
100&1.15&1.5&3&1681857&2009251&1595638\\  
500&2.75&1.5&3&35991375&44708693&30626456\\
   \midrule
100&1.2&1&3&1873354&2225262&1778719 \\ 
500&2.775&1&3&37496379&47889891&32178178\\
   \bottomrule
\end{tabular}
\end{table}

\begin{table}[h]
\centering
\caption{\label{tab:sims4mise} MISE for the global, local, and tessellated models, averaged over the 100 simulated point patterns generated assuming intensity \eqref{eq:syst3}.}
\begin{tabular}{c|cccc|rrr}
  \toprule
    & \multicolumn{4}{c|}{Parameters} &  \multicolumn{3}{c}{MISE} \\
      \midrule
   $\mathbb{E}[N]$ & $\beta_0$ & $\beta_1$ & $\gamma_0$  & $\gamma_1$ & global  & local & tessellated   \\ 
  \midrule
  100 &2.5 & 0.50& 4.95& 0.75 &13087 &19816&1608\\
  500 & 3.40&0.50 &6.64 & 0.75 &409464 &457716&45097\\
   \bottomrule
\end{tabular}
\end{table}

\section{Applications}\label{sec:appl} 
This section applies the tessellated model to the two examples introduced in Section \ref{sec:examples}.

\subsection{Gordon Square data}

Consider now the point pattern of the Gordon Square example. We fit the following tessellated model
\begin{equation}
    \label{eq:tess_gordon}
    \hat{\lambda}_{gordon}(u) = \exp\{\hat{\beta}_0 + \hat{\gamma}_0W(u)\},
\end{equation}
which basically consists of the estimation of two constant intensities in the identified tessellation which resulted having only two tiles. 
Table  \ref{tab:AIC_gordon} compares the goodness-of-fit of the tessellated model \eqref{eq:tess_gordon} and its global and local counterparts, $$\hat{\lambda}_{gordon}(u) = \exp\{\hat{\theta}_0\} \qquad \text{and} \qquad  \hat{\lambda}_{gordon}(u) = \exp\{\hat{\theta}_0(u)\}.$$

\begin{table}[h]
    \caption{AIC and MISE of the models fitted to the Gordon Square data.}
    \centering
    \begin{tabular}{l|rr}
        \toprule
        Model & AIC &MISE \\
        \midrule
       global &811 & 25\\
    tessellated           &721 & 22\\
      local       &- & 27\\
        \bottomrule
    \end{tabular}
    \label{tab:AIC_gordon}
\end{table}

For the global model, we compare the AIC, while for the local model, we have to resort to the MISE. For an observed point pattern, of course the true intensity in not known, so in the same spirit as the smoothed raw residuals \citep{alm1998approximation}, we compute the $\mathrm{MISE}_{\hat{\lambda}}$ as by $\frac{1}{n}\sum_{i=1}^n \int_W\big(\hat{\lambda}(u;\mathbf{x}_i) - \tilde{\lambda}(u)\big)^2du$, where $\tilde{\lambda}(u)=e(u)\sum_{i=1}^{n(\textbf{x})} \kappa(u-v_i)
\label{eq:smo1}$  is the non-parametric, kernel estimate of the fitted intensity $ \hat{\lambda}(u)$.  Here, $\kappa$ is the smoothing kernel, and $e(u)$ is the edge correction. 
The smoothing bandwidth for the kernel estimation of the raw residuals is selected by cross-validation as the value that minimises the MSE criterion defined by \cite{diggle1985kernel}, by the method of \cite{berman1989estimating}. See \cite{diggle:13} for further details.
Given the above, this formulation of the MISE favours non-parametric models like the local ones.

Table \ref{tab:AIC_gordon} reports a lower AIC of the tessellated model if compared to the global one and also the lowest MISE among the three fitted models. This proves the best fitting of the tessellated model. 
Figure    \ref{fig:gordon_est} shows $\hat{\beta}_0 + \hat{\gamma}_0W(u)$. From this, we know that the detected tessellation consists of two tiles only, basically distinguishing two main types of area, one characterised by a higher intensity of $\exp(\hat{\gamma}_0) = \exp(2.47) = 11.82$ and another characterised by a lower intensity of $\exp(\hat{\beta}_0) = \exp(-4.87) = 0.008$.

\begin{figure}[h]
    \centering
\vspace{-1cm}
\includegraphics[width=.5\textwidth]{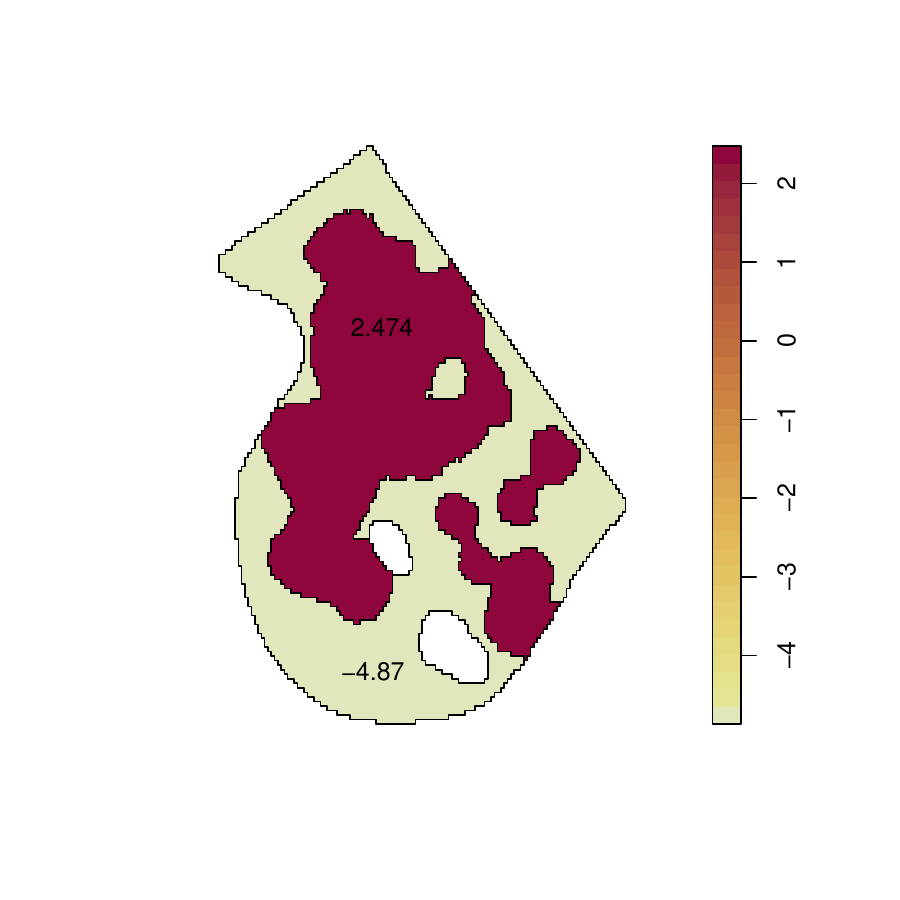}
    \caption{Spatially varying intercept estimated from model     \eqref{eq:tess_gordon} fitted to the Gordon Sqaure data.}
    \label{fig:gordon_est}
\end{figure}


Table \ref{tab:1_gordon} provides further information on the estimates, such as their significance and confidence intervals.

\begin{table}[h]
    \centering
    \caption{Estimated parameters of model \eqref{eq:tess_gordon}.}
    \begin{tabular}{llrrrrrc}
        \toprule
      Variable & Parameter  & Estimate & S.E. & CI95.lo & CI95.hi & Ztest & Zval \\
        \midrule
        Intercept  & $\hat{\beta}_0$  &  -4.87  & 0.33  &  -5.52  &  -4.22  & *** &  -14.61  \\
                $W(u)$  &  $\hat{\gamma}_0$     & 2.47  & 0.35  & 1.79  & 3.16  & *** & 7.08  \\
        \bottomrule
    \end{tabular}
\label{tab:1_gordon}
\end{table}

Moreover, we compared the feature detection with the results obtained from the algorithm of \cite{Byers1998}. That method assumes only two clusters, feature and noise, even though the algorithm could be extended to consider multiple features in a fractal framework or with just different intensities. Figure \ref{fig:class} shows the classification of the points into clutter and features (black and red points, respectively). The tessellation identified by our method is superimposed to the graphs even though it did not have a role in the feature detection with \cite{Byers1998}'s method. As regards the classification obtained from the tessellated model (left panel), the points are just classified according to their place of occurrence with respect to the identified tiles. Of course, the points belonging to the central tile with the highest intensity are classified as features, and the rest are classified as clutter. For \cite{Byers1998}'s method (right panel), the number of nearest neighbours to be considered in the EM algorithm based on the points' distances is 8. This has been selected through \cite{d2024advances}'s proposal, which considered the nearest neighbour K as that value which leads to the highest entropy measure, after which no further improvement is found.

\begin{figure}[h]
    \centering
    \vspace{-1cm}
\subfloat[Tessellated model]{\includegraphics[width=.5\textwidth]{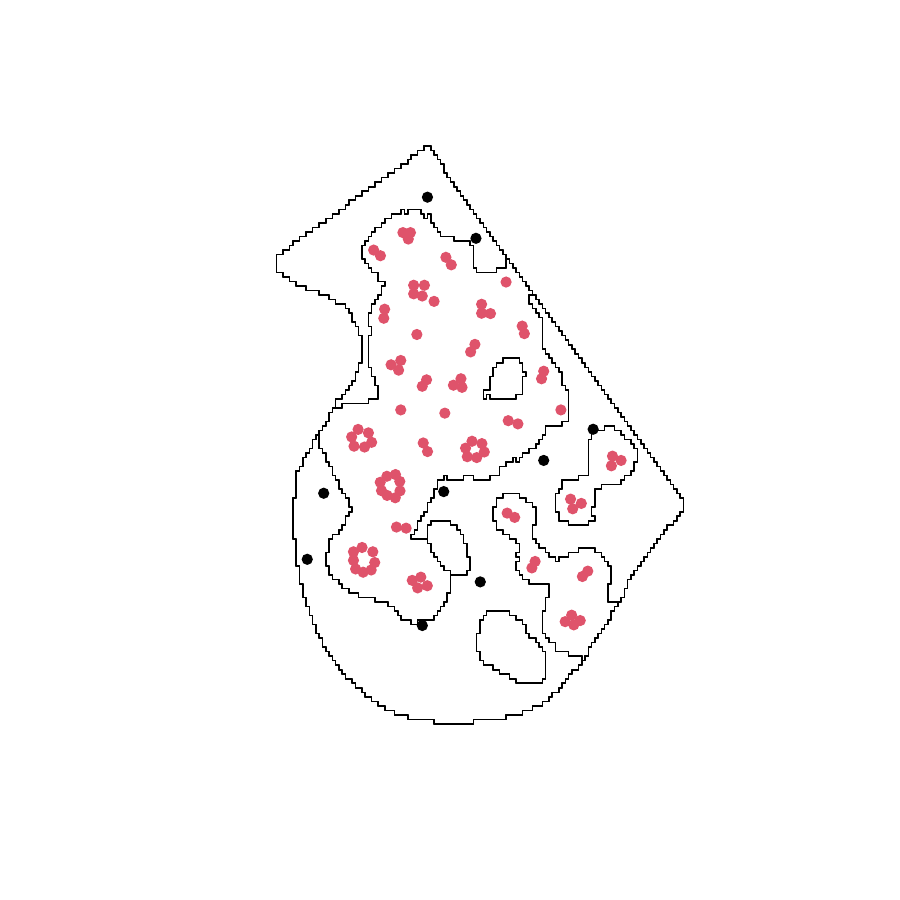}}
\subfloat[\cite{Byers1998}'s method]{\includegraphics[width=.5\textwidth]{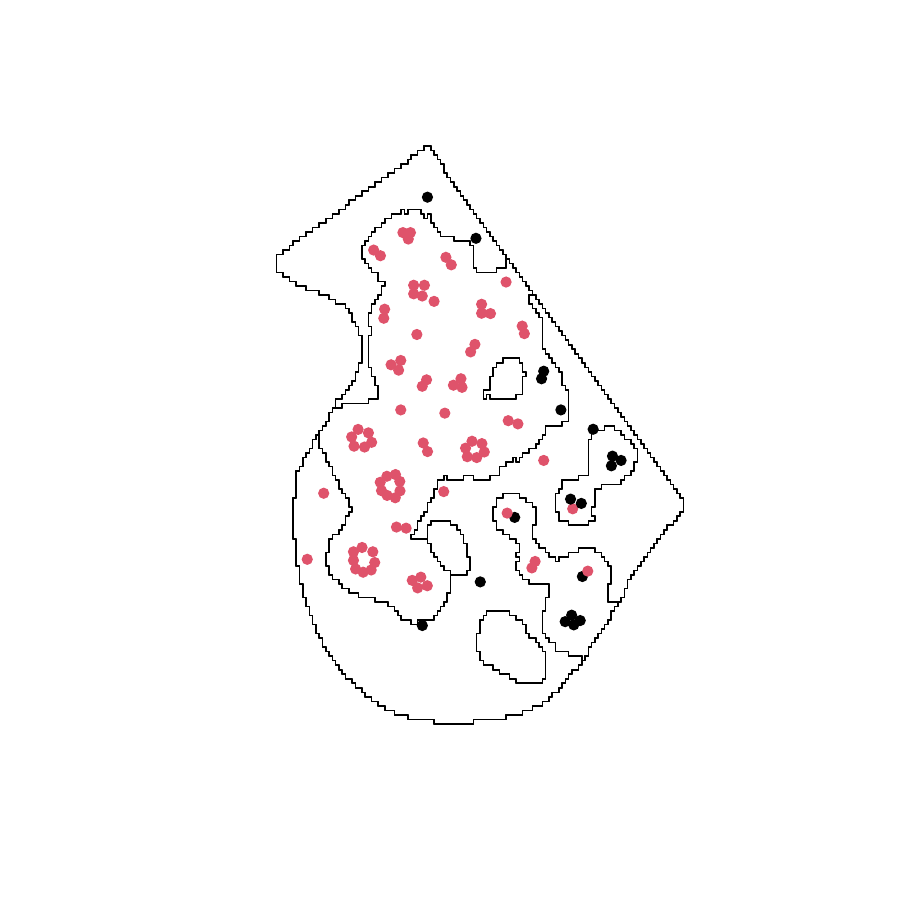}}
    \caption{Classification of the points by the tessellated regression model (a) and the feature detection algorithm of \cite{Byers1998} (b). Clutter points are in black, and feature points are in red.} 
    \label{fig:class}
\end{figure}

As evident from Figure    \ref{fig:class}, our method better succeeds in detecting all of the clusters with at least two points. On the contrary, \cite{Byers1998}'s method leaves some spatial clusters belonging to the clutter pattern. This results in the two estimated intensities being really close to each other, if considering \cite{Byers1998}'s method: 0.01 for the clutter pattern and 2.29 for the feature pattern.

\subsection{Bronze filter data}

For the bronze filter data, we fit the following tessellated model
\begin{equation}
    \label{eq:tess_bronze}
    \hat{\lambda}_{bronze}(u) = \exp\{\hat{\beta}_0 + \hat{\beta}_1 Long(u) + \hat{\gamma}_0W_{Intercept}(u) + \sum_{k=1}^4\hat{\gamma}_{1k}Long(u)W_{Long,k}(u)\}.
\end{equation}

In this case, the tessellation obtained for the intercept has only two tiles, while the tessellation for the x-coordinate covariate has five.
Table     \ref{tab:AIC_bronze} reports the AIC and MISE of the models fitted to the bronze filter data, proving also in this case the better goodness-of-fit of the tessellated model over the other competitors.
\begin{table}[h]
    \caption{AIC and MISE of the models fitted to the bronze filter data.}
    \centering
    \begin{tabular}{l|rr}
        \toprule
        Model & AIC& MISE \\
        \midrule
       global &-1117 & 140\\
    tessellated           &-1119 &101\\
      local       &- &119 \\
        \bottomrule
    \end{tabular}
    \label{tab:AIC_bronze}
\end{table}

Figure  \ref{fig:bronze_est} show the estimated coefficients, indicating a decrease in the intensity along the x-axis and a peak in the x-coordinate effect in the middle of the region.
\begin{figure}[h]
    \centering
\vspace{-.75cm}    \includegraphics[width=\textwidth]{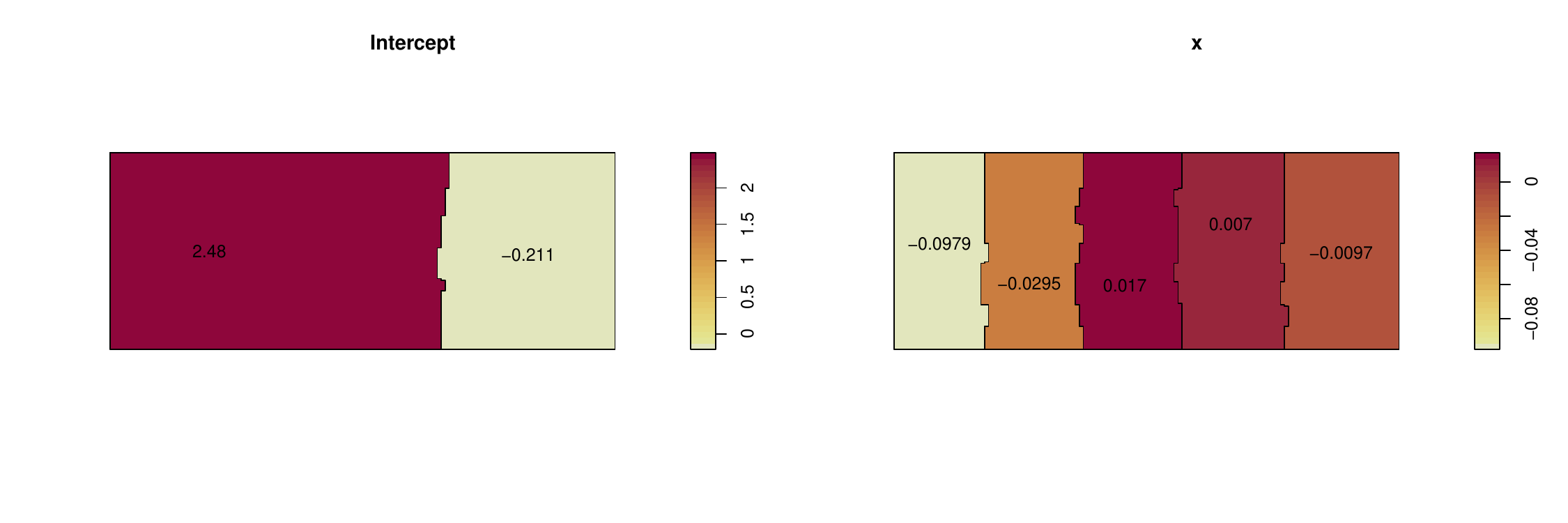}
\vspace{-2cm}
    \caption{Spatially varying coefficients of model \eqref{eq:tess_bronze} fitted to the Bronze filter data. }
    \label{fig:bronze_est}
\end{figure}

Note that even though the estimates did not resulted significant (see Table \ref{tab:1_bronze}), the tessellated model still provides better prediction if compared to the competitor models. Furthermore, the Likelihood Ration Test (LRT) on the $\boldsymbol{\gamma}$ parameters showed a p-value of 0.03, confirming the necessity to include region-wise effects in the model.

\begin{table}[h]
    \centering
    \caption{Estimated parameters of model \eqref{eq:tess_bronze}.}
    \begin{tabular}{llrrrrrc}
        \toprule
      Variable & Parameter  & Estimate & S.E. & CI95.lo & CI95.hi & Ztest & Zval \\
        \midrule
        Intercept  & $\hat{\beta}_0$  &  2.48  & 0.12  &  2.25  &  2.71  & *** &  20.95  \\
        $Long(u)$  &  $\hat{\beta}_1$     & -0.21  & 0.22  & -0.65  & 0.22  &  & -0.95  \\
        $W_{Intercept}(u)$   & $\hat{\gamma}_0$      & -0.10  & 0.07  & -0.23  & 0.03  &  &  -1.46  \\
        $W_{Long,1}(u)$ &  $\hat{\gamma}_{11}$   & -0.03  & 0.05  & -0.13  &  0.07  &  &  -0.57  \\
        $W_{Long,2}(u)$& $\hat{\gamma}_{12}$&  0.02  & 0.06  & -0.09  &  0.13  &  &  0.30  \\
       $W_{Long,3}(u)$& $\hat{\gamma}_{13}$ &  0.01  & 0.06  & -0.11  &  0.12  &  &   0.12  \\
      $W_{Long,4}(u)$& $\hat{\gamma}_{14}$ &  -0.01  & 0.06  & -0.13  &  0.11  &  &   -0.16  \\
        \bottomrule
    \end{tabular}
\label{tab:1_bronze}
\end{table}

\section{Analysis of the Greek seismicity}\label{sec:greece}

For the Greek data, we fit the following tessellated model
\begin{equation}
 \hat{\lambda}_{greece}(u) = \exp\{ \hat{\beta}_0 + \hat{\beta}_1D_p(u) + \hat{\beta}_2D_f(u) +\sum_{k=1}^4\hat{\gamma}_{0k}W_{Intercept,k}(u) + \hat{\gamma}_1 D_p(u) W_{D_p}(u)    + \hat{\gamma}_2 D_f(u) W_{D_f}(u)  \}   
 \label{eq:non_seg}
\end{equation}
and compare it to the global and local one in Table \ref{tab:AIC_seismic}.

\begin{table}[h]
    \caption{AIC and MISE of the models fitted to the Greek seismicity data.}
    \centering
\begin{tabular}{l|rr}
        \toprule
        Model & AIC &MISE \\
        \midrule
       global &-4485 &431123\\
    tessellated           & -5093&425597 \\
      local       &- &403759 \\
        \bottomrule
    \end{tabular}
    \label{tab:AIC_seismic}
\end{table}

The tessellated model has a lower AIC than the global one.
Fitting the (non-tessellated) global model, the parameter estimates $\hat{\boldsymbol{\theta}} = \{ 3.96, -0.90, -0.91 \}$ are obtained, all significantly different from zero. As expected, the distance from the seismic sources decreases the intensity of the earthquakes. We further aim to compare the tessellated model with the global one
 by conducting a formal test to assess the significance of $\boldsymbol{\gamma}$. The LRT for $\boldsymbol{\gamma}$ shows a zero \(p\)-value, indicating that the tessellated model should be preferred over the global one.

Then, from Table \ref{tab:AIC_seismic}, we know that the lowest MISE is obtained by the local model, meaning that this provides better goodness-of-fit. However, we decided to proceed with the interpretation of the tessellated model, which further provides clear-cut spatial changes effects of the model parameters. Indeed, the regions identified may reveal similar characteristics of the underlying process that generated the events in those regions.

The estimated parameters of model \eqref{eq:non_seg} are reported in Table \ref{tab:1} and shown in Figure \ref{fig:5b}.

\begin{table}[h]
    \centering
    \caption{Estimated parameters of model  \eqref{eq:non_seg}.}
    \begin{tabular}{llrrrrrc}
        \toprule
      Variable & Parameter  & Estimate & S.E. & CI95.lo & CI95.hi & Ztest & Zval \\
        \midrule
                Intercept  & ${\beta}_0$  &  5.09  & 0.07  &  4.95  &  5.23  & *** &  70.00  \\
       $W_{Intercept,1}(u)$  &  ${\gamma}_{01}$  & -1.01  & 0.08  & -1.17  & -0.86  & *** & -12.96  \\
      $W_{Intercept,2}(u)$  &  ${\gamma}_{02}$  & -1.24  & 0.09  & -1.42  & -1.07  & *** & -14.10  \\
        $W_{Intercept,3}(u)$  &  ${\gamma}_{03}$  & -2.03  & 0.14  & -2.31  & -1.75  & *** & -14.12  \\
        $W_{Intercept,4}(u)$  &  ${\gamma}_{04}$  & -3.40  & 0.41  & -4.21  & -2.59  & *** & -8.23  \\
         $W_{Intercept,5}(u)$  &  ${\gamma}_{05}$  & -2.25  & 0.29  & -2.81  & -1.69  & *** & -7.88  \\
       $D_f(u)$  &  ${\beta}_1$  & -1.44  & 0.11  & -1.65  & -1.23  & *** & -13.26  \\
        $D_p(u)$  &  ${\beta}_2$  & -0.98  & 0.07  & -1.12  & -0.84  & *** & -13.88  \\
        $D_f(u)W_{D_f}(u)$  & ${\gamma}_2$  & -2.34  & 0.28  & -2.90  & -1.78  & *** & -8.21  \\
        $D_p(u)W_{D_p}(u)$  & ${\gamma}_1$  &  0.79  & 0.10  &  0.60  &  0.98  & *** &  8.25  \\
        \bottomrule
    \end{tabular}
\label{tab:1}
\end{table}

In particular, Figure \ref{fig:5b} informs us of the spatial displacement of the estimated coefficients, aiding the interpretation. The spatially varying intercept shows regions at the west and south boundaries where the intensity is significantly higher with respect to the other identified regions. This high intensity may be attributable to some specific characteristics of those regions, not taken into account by the model. Then, the intensity gradually decreases along the northeast part of the region. However, the most crucial information that a local model would not be able to provide is the clear separation of the effects of the spatial covariates. Indeed the distance from the plate boundary seems to negatively affect the intensity of the seismic process with different magnitudes into the two identified tiles/subregions. Most interestingly, the distance from the faults has a negative effect on the process intensity in the majority of the region, as expected. The only expectation is in the northeast region where the effect becomes positive. This specific tile identifies a region where few earthquakes occurred in the presence of many faults. This finding could lead to a deeper study of the seismic phenomenon in the specific region.

\begin{figure}[h]
	\centering
\subfloat[Intercept]{\includegraphics[width=.33\textwidth]{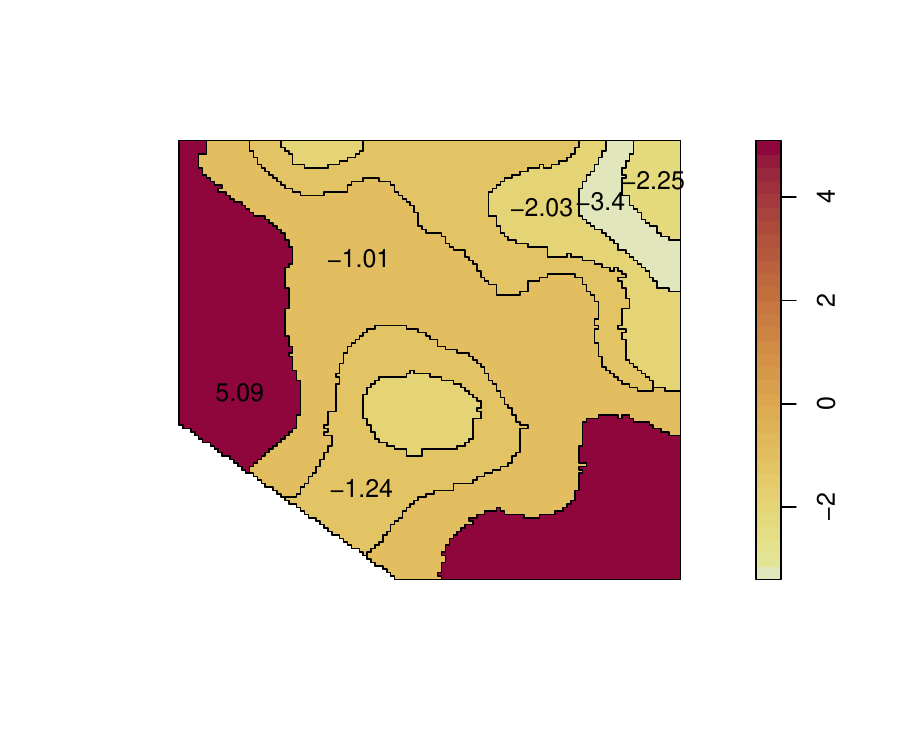}}
\subfloat[Distance from the plate boundary]{\includegraphics[width=.33\textwidth]{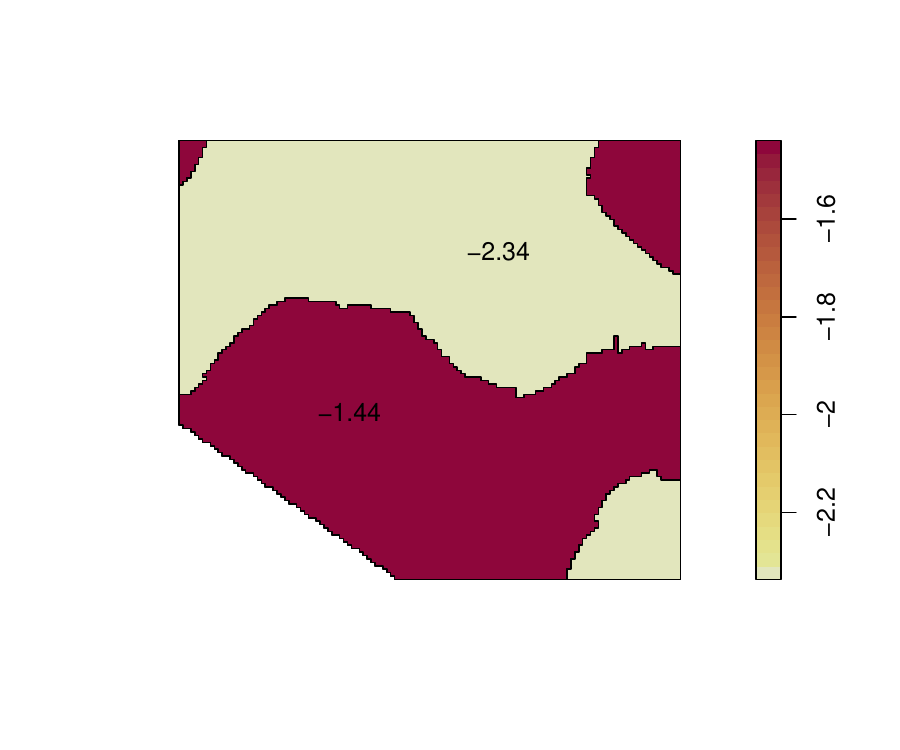}}
\subfloat[Distance from the faults]{\includegraphics[width=.33\textwidth]{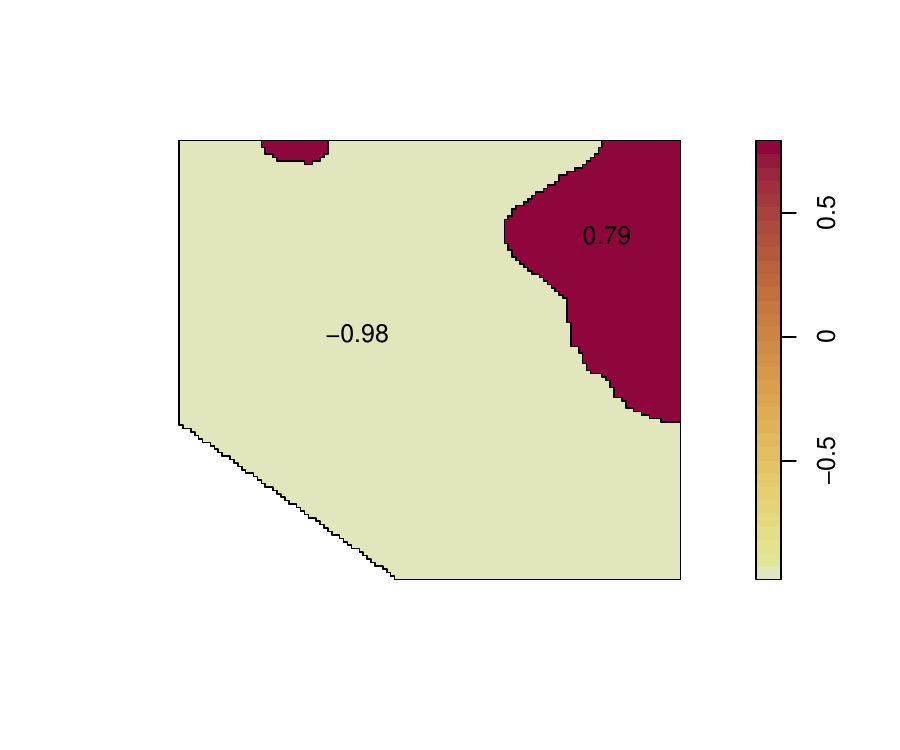}}
	\caption{Spatially varying coefficients of model  \eqref{eq:non_seg} fitted to the Greek seismicity data.}
	\label{fig:5b}
\end{figure}

\section{Conclusions}\label{sec:concl}
This work has formalised and proposed a method to fit the novel tessellated spatial log-linear Poisson point process model, which is particularly useful in the local estimation context. The motivation of the work came from the necessity to obtain clear-cut regions where the parameters of the spatial covariates in the spatial Poisson point process model change, in the same philosophy as the models with spatial varying parameters, but with the advantage of sticking to the realm of models with fixed effects parameters.

Our approach is based on a two-step procedure: tessellation identification and tessellated model fitting. For the first step, a segmentation algorithm is applied to the geographically weighted estimates of an assumed local model.  We decided to resort to the SLIC algorithm for this first step, together with a cluster analysis to group superpixels with similar value, to avoid oversegmentation. The second step is straightforwardly achieved by treating the tessellation obtained at the first step as a categorical spatial variable, encoded in some dummy variables indicating which tile the point of the pattern belongs to. This allows to obtain region-wise effects of the covariates within the framework of the most classical Poisson log-linear point process models.

The main advantage of this proposal is the ability to retain both interpretation and standard statistical procedures while also achieving a higher degree of flexibility without specifically resorting to non-parametric approaches. 

It is noteworthy to highlight that this proposal for the tessellation identification solves a common problem in \lq \lq segmented'' regression models, that is, the estimation of the number of changepoints and their position. Here the changepoints' counterparts are the tiles of the tessellations, and the SLIC algorithm, with the subsequent cluster analysis, already identifies both the number and the position of tiles prior to the actual model fitting.

With simulations, we have explored the performance of the methodology finding out, as expected, that the performance strongly relies on the number of points in the pattern but, most importantly, on the magnitude of the change in the parameters. In general, the tessellated model provides better goodness-of-fit when the effects of real-valued covariates on the process intensity are actually region-wise.
The applications to three real data problems have illustrated potential employments of the proposed methodology. These have ranged from the most simple case of only space-varying homogeneous intensity to the most complex case of multiple external real-valued spatial covariates. We ague that the former case can serve as a feature detection/clutter removal algorithm, with the further advantage, with respect to standard methodologies, of not having to assume the number of features. Furthermore, the shape of the features does not represent a problem, and it is strongly dependent on the geographically weighted estimates obtained from the local model fitted in advance.

The major limitation of our proposal is the dependence on the geographically weighted regression parameters and on the fitting of a local model, which, in turn, could be computationally demanding when dealing with big data patterns. We, however, believe this is necessary in order to fully exploit the gradual estimated changes provided by the local models and to fully exploit standard inferential tools of Poisson likelihood.

This work could be enhanced in many ways. First, other segmentation algorithms could be employed to improve the overall performance and robustness of the results.
Secondly, inference on the tessellation identification could be explored in order to provide an uncertainty measure.

We believe the formalisation of this model opens the path for many future developments and applications to various contexts. First, the same reasoning is straightforwardly applicable to the geostatistical context, where the response of the model is not the point pattern intensity but rather a proper observed response variable.
Secondly, regarding the model formalisation, another option could be to treat the belonging to the specific tile as a random effect rather than a fixed effect through the tessellation. This would imply the specification of a multitype point process model, which we decided not to address in the current work.
Of course, the extension to the spatio-temporal context would represent a relevant development, conceptually following the work carried out in the R (\cite{R}) package \texttt{stopp} (\cite{d2024stopp}). 
 Finally, the methodology could be extended to a more complex model whose local version is already defined, like the log-Gaussian Cox processes (LGCPs). Indeed, \cite{dangelo2021locall,d2022locally} fitted also local LGCPs to earthquake data, finding that the seismic phenomenon displayed a spatially varying variance as well as scale parameters, meaning that points tend to cluster in smaller clusters in specific regions of the analysed area. Segmenting such local effects is of course possible with the SLIC algorithm, but what remain for future work would be the formalisation of a tessellated LGCP, which would imply a region-wise covariance function.

\section*{Fundings}
The research work was supported by the Targeted Research Funds 2025 (FFR 2025) of the University of Palermo, GRINS PE00000018 – CUP  C93C22005270001 from the European Union -  NextGenerationEU, and PRIN 2022: Spatio-temporal Functional Marked Point Processes for
probabilistic forecasting of earthquakes 2022BN7CJP P. I. Giada Adelfio. CUP B53C24006340006.

\bibliography{BBB}

\begin{thebibliography}{}

\bibitem [\protect \citeauthoryear {%
Achanta%
\ \protect \BOthers {.}}{%
Achanta%
\ \protect \BOthers {.}}{%
{\protect \APACyear {2012}}%
}]{%
Achanta2012}
\APACinsertmetastar {%
Achanta2012}%
\begin{APACrefauthors}%
Achanta, R.%
, Shaji, A.%
, Smith, K.%
, Lucchi, A.%
, Fua, P.%
\BCBL {}\ \BBA {} S{\"u}sstrunk, S.%
\end{APACrefauthors}%
\unskip\
\newblock
\APACrefYearMonthDay{2012}{}{}.
\newblock
{\BBOQ}\APACrefatitle {SLIC Superpixels Compared to State-of-the-Art Superpixel
  Methods} {Slic superpixels compared to state-of-the-art superpixel
  methods}.{\BBCQ}
\newblock
\APACjournalVolNumPages{IEEE Transactions on Pattern Analysis and Machine
  Intelligence}{34}{11}{2274--2282}.
\newblock
\begin{APACrefDOI} 10.1109/TPAMI.2012.120 \end{APACrefDOI}
\PrintBackRefs{\CurrentBib}

\bibitem [\protect \citeauthoryear {%
Allard%
\ \BBA {} Fraley%
}{%
Allard%
\ \BBA {} Fraley%
}{%
{\protect \APACyear {1997}}%
}]{%
allard1997nonparametric}
\APACinsertmetastar {%
allard1997nonparametric}%
\begin{APACrefauthors}%
Allard, D.%
\BCBT {}\ \BBA {} Fraley, C.%
\end{APACrefauthors}%
\unskip\
\newblock
\APACrefYearMonthDay{1997}{}{}.
\newblock
{\BBOQ}\APACrefatitle {Nonparametric maximum likelihood estimation of features
  in spatial point processes using Voronoi tessellation} {Nonparametric maximum
  likelihood estimation of features in spatial point processes using voronoi
  tessellation}.{\BBCQ}
\newblock
\APACjournalVolNumPages{Journal of the American statistical
  Association}{92}{440}{1485--1493}.
\PrintBackRefs{\CurrentBib}

\bibitem [\protect \citeauthoryear {%
Alm%
}{%
Alm%
}{%
{\protect \APACyear {1998}}%
}]{%
alm1998approximation}
\APACinsertmetastar {%
alm1998approximation}%
\begin{APACrefauthors}%
Alm, S\BPBI E.%
\end{APACrefauthors}%
\unskip\
\newblock
\APACrefYearMonthDay{1998}{}{}.
\newblock
{\BBOQ}\APACrefatitle {Approximation and simulation of the distributions of
  scan statistics for Poisson processes in higher dimensions} {Approximation
  and simulation of the distributions of scan statistics for poisson processes
  in higher dimensions}.{\BBCQ}
\newblock
\APACjournalVolNumPages{Extremes}{1}{1}{111--126}.
\PrintBackRefs{\CurrentBib}

\bibitem [\protect \citeauthoryear {%
Altieri%
, Scott%
, Cocchi%
\BCBL {}\ \BBA {} Illian%
}{%
Altieri%
\ \protect \BOthers {.}}{%
{\protect \APACyear {2015}}%
}]{%
altieri2015changepoint}
\APACinsertmetastar {%
altieri2015changepoint}%
\begin{APACrefauthors}%
Altieri, L.%
, Scott, E\BPBI M.%
, Cocchi, D.%
\BCBL {}\ \BBA {} Illian, J\BPBI B.%
\end{APACrefauthors}%
\unskip\
\newblock
\APACrefYearMonthDay{2015}{}{}.
\newblock
{\BBOQ}\APACrefatitle {A changepoint analysis of spatio-temporal point
  processes} {A changepoint analysis of spatio-temporal point
  processes}.{\BBCQ}
\newblock
\APACjournalVolNumPages{Spatial Statistics}{14}{}{197--207}.
\PrintBackRefs{\CurrentBib}

\bibitem [\protect \citeauthoryear {%
Baddeley%
}{%
Baddeley%
}{%
{\protect \APACyear {2017}}%
}]{%
baddeley:2017local}
\APACinsertmetastar {%
baddeley:2017local}%
\begin{APACrefauthors}%
Baddeley, A.%
\end{APACrefauthors}%
\unskip\
\newblock
\APACrefYearMonthDay{2017}{}{}.
\newblock
{\BBOQ}\APACrefatitle {Local composite likelihood for spatial point processes}
  {Local composite likelihood for spatial point processes}.{\BBCQ}
\newblock
\APACjournalVolNumPages{Spatial Statistics}{22}{}{261--295}.
\PrintBackRefs{\CurrentBib}

\bibitem [\protect \citeauthoryear {%
Baddeley%
, Coeurjolly%
, Rubak%
\BCBL {}\ \BBA {} Waagepetersen%
}{%
Baddeley%
\ \protect \BOthers {.}}{%
{\protect \APACyear {2014}}%
}]{%
baddeley2014logistic}
\APACinsertmetastar {%
baddeley2014logistic}%
\begin{APACrefauthors}%
Baddeley, A.%
, Coeurjolly, J\BHBI F.%
, Rubak, E.%
\BCBL {}\ \BBA {} Waagepetersen, R.%
\end{APACrefauthors}%
\unskip\
\newblock
\APACrefYearMonthDay{2014}{}{}.
\newblock
{\BBOQ}\APACrefatitle {Logistic regression for spatial Gibbs point processes}
  {Logistic regression for spatial gibbs point processes}.{\BBCQ}
\newblock
\APACjournalVolNumPages{Biometrika}{101}{2}{377--392}.
\PrintBackRefs{\CurrentBib}

\bibitem [\protect \citeauthoryear {%
Baddeley%
, Rubak%
\BCBL {}\ \BBA {} Turner%
}{%
Baddeley%
\ \protect \BOthers {.}}{%
{\protect \APACyear {2015}}%
}]{%
baddeley:rubak:tuner:15}
\APACinsertmetastar {%
baddeley:rubak:tuner:15}%
\begin{APACrefauthors}%
Baddeley, A.%
, Rubak, E.%
\BCBL {}\ \BBA {} Turner, R.%
\end{APACrefauthors}%
\unskip\
\newblock
\APACrefYear{2015}.
\newblock
\APACrefbtitle {Spatial Point Patterns: Methodology and Applications with R}
  {Spatial point patterns: Methodology and applications with r}.
\newblock
\APACaddressPublisher{}{London: Chapman and Hall/CRC Press}.
\PrintBackRefs{\CurrentBib}

\bibitem [\protect \citeauthoryear {%
Baddeley%
\ \BBA {} Turner%
}{%
Baddeley%
\ \BBA {} Turner%
}{%
{\protect \APACyear {2005}}%
}]{%
spat}
\APACinsertmetastar {%
spat}%
\begin{APACrefauthors}%
Baddeley, A.%
\BCBT {}\ \BBA {} Turner, R.%
\end{APACrefauthors}%
\unskip\
\newblock
\APACrefYearMonthDay{2005}{}{}.
\newblock
{\BBOQ}\APACrefatitle {{spatstat}: An {R} Package for Analyzing Spatial Point
  Patterns} {{spatstat}: An {R} package for analyzing spatial point
  patterns}.{\BBCQ}
\newblock
\APACjournalVolNumPages{Journal of Statistical Software}{12}{6}{1--42}.
\newblock
\begin{APACrefURL} \url{http://www.jstatsoft.org/v12/i06/} \end{APACrefURL}
\PrintBackRefs{\CurrentBib}

\bibitem [\protect \citeauthoryear {%
Baddeley%
, Turner%
, Mateu%
\BCBL {}\ \BBA {} Bevan%
}{%
Baddeley%
\ \protect \BOthers {.}}{%
{\protect \APACyear {2013}}%
}]{%
baddeley2013hybrids}
\APACinsertmetastar {%
baddeley2013hybrids}%
\begin{APACrefauthors}%
Baddeley, A.%
, Turner, R.%
, Mateu, J.%
\BCBL {}\ \BBA {} Bevan, A.%
\end{APACrefauthors}%
\unskip\
\newblock
\APACrefYearMonthDay{2013}{}{}.
\newblock
{\BBOQ}\APACrefatitle {Hybrids of Gibbs point process models and their
  implementation} {Hybrids of gibbs point process models and their
  implementation}.{\BBCQ}
\newblock
\APACjournalVolNumPages{Journal of Statistical Software}{55}{}{1--43}.
\PrintBackRefs{\CurrentBib}

\bibitem [\protect \citeauthoryear {%
Beckman%
\ \BBA {} Cook%
}{%
Beckman%
\ \BBA {} Cook%
}{%
{\protect \APACyear {1979}}%
}]{%
beckman1979testing}
\APACinsertmetastar {%
beckman1979testing}%
\begin{APACrefauthors}%
Beckman, R.%
\BCBT {}\ \BBA {} Cook, R.%
\end{APACrefauthors}%
\unskip\
\newblock
\APACrefYearMonthDay{1979}{}{}.
\newblock
{\BBOQ}\APACrefatitle {Testing for two-phase regressions} {Testing for
  two-phase regressions}.{\BBCQ}
\newblock
\APACjournalVolNumPages{Technometrics}{21}{1}{65--69}.
\PrintBackRefs{\CurrentBib}

\bibitem [\protect \citeauthoryear {%
Berman%
\ \BBA {} Diggle%
}{%
Berman%
\ \BBA {} Diggle%
}{%
{\protect \APACyear {1989}}%
}]{%
berman1989estimating}
\APACinsertmetastar {%
berman1989estimating}%
\begin{APACrefauthors}%
Berman, M.%
\BCBT {}\ \BBA {} Diggle, P.%
\end{APACrefauthors}%
\unskip\
\newblock
\APACrefYearMonthDay{1989}{}{}.
\newblock
{\BBOQ}\APACrefatitle {Estimating weighted integrals of the second-order
  intensity of a spatial point process} {Estimating weighted integrals of the
  second-order intensity of a spatial point process}.{\BBCQ}
\newblock
\APACjournalVolNumPages{Journal of the Royal Statistical Society: Series B
  (Methodological)}{51}{1}{81--92}.
\PrintBackRefs{\CurrentBib}

\bibitem [\protect \citeauthoryear {%
Berman%
\ \BBA {} Turner%
}{%
Berman%
\ \BBA {} Turner%
}{%
{\protect \APACyear {1992}}%
}]{%
berman1992approximating}
\APACinsertmetastar {%
berman1992approximating}%
\begin{APACrefauthors}%
Berman, M.%
\BCBT {}\ \BBA {} Turner, T\BPBI R.%
\end{APACrefauthors}%
\unskip\
\newblock
\APACrefYearMonthDay{1992}{}{}.
\newblock
{\BBOQ}\APACrefatitle {Approximating point process likelihoods with GLIM}
  {Approximating point process likelihoods with glim}.{\BBCQ}
\newblock
\APACjournalVolNumPages{Journal of the Royal Statistical Society: Series C
  (Applied Statistics)}{41}{1}{31--38}.
\PrintBackRefs{\CurrentBib}

\bibitem [\protect \citeauthoryear {%
Bernhardt%
, Meyer-Olbersleben%
\BCBL {}\ \BBA {} Kieback%
}{%
Bernhardt%
\ \protect \BOthers {.}}{%
{\protect \APACyear {1997}}%
}]{%
bernhardt1997fundamental}
\APACinsertmetastar {%
bernhardt1997fundamental}%
\begin{APACrefauthors}%
Bernhardt, R.%
, Meyer-Olbersleben, F.%
\BCBL {}\ \BBA {} Kieback, B.%
\end{APACrefauthors}%
\unskip\
\newblock
\APACrefYearMonthDay{1997}{}{}.
\newblock
{\BBOQ}\APACrefatitle {Fundamental investigation on the preparation of gradient
  structures by sedimentation of different powder fractions under gravity}
  {Fundamental investigation on the preparation of gradient structures by
  sedimentation of different powder fractions under gravity}.{\BBCQ}
\newblock
\BIn{} \APACrefbtitle {Proc. of the 4th Int. Conf. on Composite Engineering}
  {Proc. of the 4th int. conf. on composite engineering}\ (\BPGS\ 147--148).
\PrintBackRefs{\CurrentBib}

\bibitem [\protect \citeauthoryear {%
Byers%
\ \BBA {} Raftery%
}{%
Byers%
\ \BBA {} Raftery%
}{%
{\protect \APACyear {1998}}%
}]{%
Byers1998}
\APACinsertmetastar {%
Byers1998}%
\begin{APACrefauthors}%
Byers, S.%
\BCBT {}\ \BBA {} Raftery, A\BPBI E.%
\end{APACrefauthors}%
\unskip\
\newblock
\APACrefYearMonthDay{1998}{}{}.
\newblock
{\BBOQ}\APACrefatitle {Nearest-Neighbor Clutter Removal for Estimating Features
  in Spatial Point Processes} {Nearest-neighbor clutter removal for estimating
  features in spatial point processes}.{\BBCQ}
\newblock
\APACjournalVolNumPages{Journal of the American Statistical
  Association}{93}{442}{577-584}.
\newblock
\begin{APACrefURL} \url{http://www.jstor.org/stable/2670109} \end{APACrefURL}
\PrintBackRefs{\CurrentBib}

\bibitem [\protect \citeauthoryear {%
Daley%
\ \BBA {} Vere-Jones%
}{%
Daley%
\ \BBA {} Vere-Jones%
}{%
{\protect \APACyear {2007}}%
}]{%
daley:vere-jones:08}
\APACinsertmetastar {%
daley:vere-jones:08}%
\begin{APACrefauthors}%
Daley, D\BPBI J.%
\BCBT {}\ \BBA {} Vere-Jones, D.%
\end{APACrefauthors}%
\unskip\
\newblock
\APACrefYear{2007}.
\newblock
\APACrefbtitle {An Introduction to the Theory of Point Processes. Volume II:
  General Theory and Structure} {An introduction to the theory of point
  processes. volume ii: General theory and structure}\ (\PrintOrdinal{Second}\
  \BEd).
\newblock
\APACaddressPublisher{}{Springer-Verlag, New York}.
\PrintBackRefs{\CurrentBib}

\bibitem [\protect \citeauthoryear {%
D'Angelo%
\ \BBA {} Adelfio%
}{%
D'Angelo%
\ \BBA {} Adelfio%
}{%
{\protect \APACyear {2024}}%
}]{%
d2024minimum}
\APACinsertmetastar {%
d2024minimum}%
\begin{APACrefauthors}%
D'Angelo, N.%
\BCBT {}\ \BBA {} Adelfio, G.%
\end{APACrefauthors}%
\unskip\
\newblock
\APACrefYearMonthDay{2024}{}{}.
\newblock
{\BBOQ}\APACrefatitle {Minimum Contrast for the First-Order Intensity
  Estimation of Spatial and Spatio-Temporal Point Processes} {Minimum contrast
  for the first-order intensity estimation of spatial and spatio-temporal point
  processes}.{\BBCQ}
\newblock
\APACjournalVolNumPages{Statistical Papers}{}{}{}.
\newblock
\begin{APACrefDOI} 10.1007/s00362-024-01541-5 \end{APACrefDOI}
\PrintBackRefs{\CurrentBib}

\bibitem [\protect \citeauthoryear {%
D'Angelo%
\ \BBA {} Adelfio%
}{%
D'Angelo%
\ \BBA {} Adelfio%
}{%
{\protect \APACyear {2025}}%
}]{%
d2024stopp}
\APACinsertmetastar {%
d2024stopp}%
\begin{APACrefauthors}%
D'Angelo, N.%
\BCBT {}\ \BBA {} Adelfio, G.%
\end{APACrefauthors}%
\unskip\
\newblock
\APACrefYearMonthDay{2025}{}{}.
\newblock
{\BBOQ}\APACrefatitle {stopp: An R Package for Spatio-Temporal Point Pattern
  Analysis} {stopp: An r package for spatio-temporal point pattern
  analysis}.{\BBCQ}
\newblock
\APACjournalVolNumPages{Journal of Statistical Software. To appear}{}{}{}.
\PrintBackRefs{\CurrentBib}

\bibitem [\protect \citeauthoryear {%
D'Angelo%
, Adelfio%
\BCBL {}\ \BBA {} Mateu%
}{%
D'Angelo%
\ \protect \BOthers {.}}{%
{\protect \APACyear {2023}}%
}]{%
d2022locally}
\APACinsertmetastar {%
d2022locally}%
\begin{APACrefauthors}%
D'Angelo, N.%
, Adelfio, G.%
\BCBL {}\ \BBA {} Mateu, J.%
\end{APACrefauthors}%
\unskip\
\newblock
\APACrefYearMonthDay{2023}{}{}.
\newblock
{\BBOQ}\APACrefatitle {Locally weighted minimum contrast estimation for
  spatio-temporal log-Gaussian Cox processes} {Locally weighted minimum
  contrast estimation for spatio-temporal log-gaussian cox processes}.{\BBCQ}
\newblock
\APACjournalVolNumPages{Computational Statistics \& Data
  Analysis}{180}{}{107679}.
\PrintBackRefs{\CurrentBib}

\bibitem [\protect \citeauthoryear {%
D'Angelo%
, Albano%
, Gilardi%
\BCBL {}\ \BBA {} Adelfio%
}{%
D'Angelo%
\ \protect \BOthers {.}}{%
{\protect \APACyear {2025}}%
}]{%
d2025non}
\APACinsertmetastar {%
d2025non}%
\begin{APACrefauthors}%
D'Angelo, N.%
, Albano, A.%
, Gilardi, A.%
\BCBL {}\ \BBA {} Adelfio, G.%
\end{APACrefauthors}%
\unskip\
\newblock
\APACrefYearMonthDay{2025}{}{}.
\newblock
{\BBOQ}\APACrefatitle {Non-separable spatio-temporal Poisson point process
  models for fire occurrences} {Non-separable spatio-temporal poisson point
  process models for fire occurrences}.{\BBCQ}
\newblock
\APACjournalVolNumPages{Environmental and Ecological
  Statistics}{32}{}{347--381}.
\newblock
\begin{APACrefDOI} 10.1007/s10651-025-00645-x \end{APACrefDOI}
\PrintBackRefs{\CurrentBib}

\bibitem [\protect \citeauthoryear {%
D'Angelo%
, Siino%
, D'Alessandro%
\BCBL {}\ \BBA {} Adelfio%
}{%
D'Angelo%
\ \protect \BOthers {.}}{%
{\protect \APACyear {2022}}%
}]{%
dangelo2021locall}
\APACinsertmetastar {%
dangelo2021locall}%
\begin{APACrefauthors}%
D'Angelo, N.%
, Siino, M.%
, D'Alessandro, A.%
\BCBL {}\ \BBA {} Adelfio, G.%
\end{APACrefauthors}%
\unskip\
\newblock
\APACrefYearMonthDay{2022}{}{}.
\newblock
{\BBOQ}\APACrefatitle {Local Spatial Log-Gaussian Cox Processes for seismic
  data} {Local spatial log-gaussian cox processes for seismic data}.{\BBCQ}
\newblock
\APACjournalVolNumPages{Advances in Statistical Analysis.
  \url{https://doi.org/10.1007/s10182-022-00444-w}}{}{}{}.
\PrintBackRefs{\CurrentBib}

\bibitem [\protect \citeauthoryear {%
P.~Diggle%
}{%
P.~Diggle%
}{%
{\protect \APACyear {1985}}%
}]{%
diggle1985kernel}
\APACinsertmetastar {%
diggle1985kernel}%
\begin{APACrefauthors}%
Diggle, P.%
\end{APACrefauthors}%
\unskip\
\newblock
\APACrefYearMonthDay{1985}{}{}.
\newblock
{\BBOQ}\APACrefatitle {A kernel method for smoothing point process data} {A
  kernel method for smoothing point process data}.{\BBCQ}
\newblock
\APACjournalVolNumPages{Journal of the Royal Statistical Society: Series C
  (Applied Statistics)}{34}{2}{138--147}.
\PrintBackRefs{\CurrentBib}

\bibitem [\protect \citeauthoryear {%
P\BPBI J.~Diggle%
}{%
P\BPBI J.~Diggle%
}{%
{\protect \APACyear {2013}}%
}]{%
diggle:13}
\APACinsertmetastar {%
diggle:13}%
\begin{APACrefauthors}%
Diggle, P\BPBI J.%
\end{APACrefauthors}%
\unskip\
\newblock
\APACrefYear{2013}.
\newblock
\APACrefbtitle {Statistical Analysis of Spatial and Spatio-Temporal Point
  Patterns} {Statistical analysis of spatial and spatio-temporal point
  patterns}.
\newblock
\APACaddressPublisher{}{CRC Press}.
\PrintBackRefs{\CurrentBib}

\bibitem [\protect \citeauthoryear {%
D’Angelo%
}{%
D’Angelo%
}{%
{\protect \APACyear {2024}}%
}]{%
d2024advances}
\APACinsertmetastar {%
d2024advances}%
\begin{APACrefauthors}%
D’Angelo, N.%
\end{APACrefauthors}%
\unskip\
\newblock
\APACrefYearMonthDay{2024}{}{}.
\newblock
{\BBOQ}\APACrefatitle {Advances in Kth nearest-neighbour clutter removal}
  {Advances in kth nearest-neighbour clutter removal}.{\BBCQ}
\newblock
\APACjournalVolNumPages{Environmental and Ecological
  Statistics}{31}{2}{537--554}.
\PrintBackRefs{\CurrentBib}

\bibitem [\protect \citeauthoryear {%
Fan%
, Wu%
\BCBL {}\ \BBA {} Feng%
}{%
Fan%
\ \protect \BOthers {.}}{%
{\protect \APACyear {2009}}%
}]{%
fan2009local}
\APACinsertmetastar {%
fan2009local}%
\begin{APACrefauthors}%
Fan, J.%
, Wu, Y.%
\BCBL {}\ \BBA {} Feng, Y.%
\end{APACrefauthors}%
\unskip\
\newblock
\APACrefYearMonthDay{2009}{}{}.
\newblock
{\BBOQ}\APACrefatitle {Local quasi-likelihood with a parametric guide} {Local
  quasi-likelihood with a parametric guide}.{\BBCQ}
\newblock
\APACjournalVolNumPages{Annals of statistics}{37}{6B}{4153}.
\PrintBackRefs{\CurrentBib}

\bibitem [\protect \citeauthoryear {%
Feder%
}{%
Feder%
}{%
{\protect \APACyear {1975}}%
}]{%
feder1975log}
\APACinsertmetastar {%
feder1975log}%
\begin{APACrefauthors}%
Feder, P\BPBI I.%
\end{APACrefauthors}%
\unskip\
\newblock
\APACrefYearMonthDay{1975}{}{}.
\newblock
{\BBOQ}\APACrefatitle {The log likelihood ratio in segmented regression} {The
  log likelihood ratio in segmented regression}.{\BBCQ}
\newblock
\APACjournalVolNumPages{The Annals of Statistics}{3}{1}{84--97}.
\PrintBackRefs{\CurrentBib}

\bibitem [\protect \citeauthoryear {%
Hjort%
\ \BBA {} Jones%
}{%
Hjort%
\ \BBA {} Jones%
}{%
{\protect \APACyear {1996}}%
}]{%
hjort1996locally}
\APACinsertmetastar {%
hjort1996locally}%
\begin{APACrefauthors}%
Hjort, N\BPBI L.%
\BCBT {}\ \BBA {} Jones, M\BPBI C.%
\end{APACrefauthors}%
\unskip\
\newblock
\APACrefYearMonthDay{1996}{}{}.
\newblock
{\BBOQ}\APACrefatitle {Locally parametric nonparametric density estimation}
  {Locally parametric nonparametric density estimation}.{\BBCQ}
\newblock
\APACjournalVolNumPages{The Annals of Statistics}{}{}{1619--1647}.
\PrintBackRefs{\CurrentBib}

\bibitem [\protect \citeauthoryear {%
Hollaway%
\ \BBA {} Killick%
}{%
Hollaway%
\ \BBA {} Killick%
}{%
{\protect \APACyear {2024}}%
}]{%
hollaway2024detection}
\APACinsertmetastar {%
hollaway2024detection}%
\begin{APACrefauthors}%
Hollaway, M\BPBI J.%
\BCBT {}\ \BBA {} Killick, R.%
\end{APACrefauthors}%
\unskip\
\newblock
\APACrefYearMonthDay{2024}{}{}.
\newblock
{\BBOQ}\APACrefatitle {Detection of spatiotemporal changepoints: a generalised
  additive model approach} {Detection of spatiotemporal changepoints: a
  generalised additive model approach}.{\BBCQ}
\newblock
\APACjournalVolNumPages{Statistics and Computing}{34}{5}{162}.
\PrintBackRefs{\CurrentBib}

\bibitem [\protect \citeauthoryear {%
Illian%
, Penttinen%
, Stoyan%
\BCBL {}\ \BBA {} Stoyan%
}{%
Illian%
\ \protect \BOthers {.}}{%
{\protect \APACyear {2008}}%
}]{%
illian:penttinen:stoyan:stoyan:08}
\APACinsertmetastar {%
illian:penttinen:stoyan:stoyan:08}%
\begin{APACrefauthors}%
Illian, J.%
, Penttinen, A.%
, Stoyan, H.%
\BCBL {}\ \BBA {} Stoyan, D.%
\end{APACrefauthors}%
\unskip\
\newblock
\APACrefYear{2008}.
\newblock
\APACrefbtitle {Statistical Analysis and Modelling of Spatial Point Patterns}
  {Statistical analysis and modelling of spatial point patterns}\ (\BVOL~70).
\newblock
\APACaddressPublisher{}{John Wiley \& Sons}.
\PrintBackRefs{\CurrentBib}

\bibitem [\protect \citeauthoryear {%
Lerman%
}{%
Lerman%
}{%
{\protect \APACyear {1980}}%
}]{%
lerman1980fitting}
\APACinsertmetastar {%
lerman1980fitting}%
\begin{APACrefauthors}%
Lerman, P.%
\end{APACrefauthors}%
\unskip\
\newblock
\APACrefYearMonthDay{1980}{}{}.
\newblock
{\BBOQ}\APACrefatitle {Fitting segmented regression models by grid search}
  {Fitting segmented regression models by grid search}.{\BBCQ}
\newblock
\APACjournalVolNumPages{Journal of the Royal Statistical Society: Series C
  (Applied Statistics)}{29}{1}{77--84}.
\PrintBackRefs{\CurrentBib}

\bibitem [\protect \citeauthoryear {%
Loader%
}{%
Loader%
}{%
{\protect \APACyear {1996}}%
}]{%
loader1996local}
\APACinsertmetastar {%
loader1996local}%
\begin{APACrefauthors}%
Loader, C\BPBI R.%
\end{APACrefauthors}%
\unskip\
\newblock
\APACrefYearMonthDay{1996}{}{}.
\newblock
{\BBOQ}\APACrefatitle {Local likelihood density estimation} {Local likelihood
  density estimation}.{\BBCQ}
\newblock
\APACjournalVolNumPages{The Annals of Statistics}{24}{4}{1602--1618}.
\PrintBackRefs{\CurrentBib}

\bibitem [\protect \citeauthoryear {%
Loader%
\ \protect \BOthers {.}}{%
Loader%
\ \protect \BOthers {.}}{%
{\protect \APACyear {1999}}%
}]{%
loader1999bandwidth}
\APACinsertmetastar {%
loader1999bandwidth}%
\begin{APACrefauthors}%
Loader, C\BPBI R.%
\BCBT {}\ \BOthersPeriod {.}
\end{APACrefauthors}%
\unskip\
\newblock
\APACrefYearMonthDay{1999}{}{}.
\newblock
{\BBOQ}\APACrefatitle {Bandwidth selection: classical or plug-in?} {Bandwidth
  selection: classical or plug-in?}{\BBCQ}
\newblock
\APACjournalVolNumPages{The Annals of Statistics}{27}{2}{415--438}.
\PrintBackRefs{\CurrentBib}

\bibitem [\protect \citeauthoryear {%
M{\o}ller%
, Syversveen%
\BCBL {}\ \BBA {} Waagepetersen%
}{%
M{\o}ller%
\ \protect \BOthers {.}}{%
{\protect \APACyear {1998}}%
}]{%
moller:98}
\APACinsertmetastar {%
moller:98}%
\begin{APACrefauthors}%
M{\o}ller, J.%
, Syversveen, A\BPBI R.%
\BCBL {}\ \BBA {} Waagepetersen, R\BPBI P.%
\end{APACrefauthors}%
\unskip\
\newblock
\APACrefYearMonthDay{1998}{}{}.
\newblock
{\BBOQ}\APACrefatitle {Log gaussian cox processes} {Log gaussian cox
  processes}.{\BBCQ}
\newblock
\APACjournalVolNumPages{Scandinavian Journal of Statistics}{25}{3}{451--482}.
\PrintBackRefs{\CurrentBib}

\bibitem [\protect \citeauthoryear {%
Monir~Hossain%
\ \BBA {} Lawson%
}{%
Monir~Hossain%
\ \BBA {} Lawson%
}{%
{\protect \APACyear {2006}}%
}]{%
monir2006cluster}
\APACinsertmetastar {%
monir2006cluster}%
\begin{APACrefauthors}%
Monir~Hossain, M.%
\BCBT {}\ \BBA {} Lawson, A\BPBI B.%
\end{APACrefauthors}%
\unskip\
\newblock
\APACrefYearMonthDay{2006}{}{}.
\newblock
{\BBOQ}\APACrefatitle {Cluster detection diagnostics for small area health
  data: with reference to evaluation of local likelihood models} {Cluster
  detection diagnostics for small area health data: with reference to
  evaluation of local likelihood models}.{\BBCQ}
\newblock
\APACjournalVolNumPages{Statistics in medicine}{25}{5}{771--786}.
\PrintBackRefs{\CurrentBib}

\bibitem [\protect \citeauthoryear {%
Muggeo%
}{%
Muggeo%
}{%
{\protect \APACyear {2003}}%
}]{%
muggeo:03}
\APACinsertmetastar {%
muggeo:03}%
\begin{APACrefauthors}%
Muggeo, V\BPBI M.%
\end{APACrefauthors}%
\unskip\
\newblock
\APACrefYearMonthDay{2003}{}{}.
\newblock
{\BBOQ}\APACrefatitle {Estimating regression models with unknown break-points}
  {Estimating regression models with unknown break-points}.{\BBCQ}
\newblock
\APACjournalVolNumPages{Statistics in Medicine}{22}{19}{3055--3071}.
\PrintBackRefs{\CurrentBib}

\bibitem [\protect \citeauthoryear {%
Nowosad%
\ \BBA {} Stepinski%
}{%
Nowosad%
\ \BBA {} Stepinski%
}{%
{\protect \APACyear {2022}}%
}]{%
nowosad2022extended}
\APACinsertmetastar {%
nowosad2022extended}%
\begin{APACrefauthors}%
Nowosad, J.%
\BCBT {}\ \BBA {} Stepinski, T\BPBI F.%
\end{APACrefauthors}%
\unskip\
\newblock
\APACrefYearMonthDay{2022}{}{}.
\newblock
{\BBOQ}\APACrefatitle {Extended SLIC superpixels algorithm for applications to
  non-imagery geospatial rasters} {Extended slic superpixels algorithm for
  applications to non-imagery geospatial rasters}.{\BBCQ}
\newblock
\APACjournalVolNumPages{International Journal of Applied Earth Observation and
  Geoinformation}{112}{}{102935}.
\PrintBackRefs{\CurrentBib}

\bibitem [\protect \citeauthoryear {%
Ogata%
\ \BBA {} Katsura%
}{%
Ogata%
\ \BBA {} Katsura%
}{%
{\protect \APACyear {1988}}%
}]{%
ogata1988likelihood}
\APACinsertmetastar {%
ogata1988likelihood}%
\begin{APACrefauthors}%
Ogata, Y.%
\BCBT {}\ \BBA {} Katsura, K.%
\end{APACrefauthors}%
\unskip\
\newblock
\APACrefYearMonthDay{1988}{}{}.
\newblock
{\BBOQ}\APACrefatitle {Likelihood analysis of spatial inhomogeneity for marked
  point patterns} {Likelihood analysis of spatial inhomogeneity for marked
  point patterns}.{\BBCQ}
\newblock
\APACjournalVolNumPages{Annals of the Institute of Statistical
  Mathematics}{40}{}{29--39}.
\PrintBackRefs{\CurrentBib}

\bibitem [\protect \citeauthoryear {%
Osborne%
, Foody%
\BCBL {}\ \BBA {} Su{\'a}rez-Seoane%
}{%
Osborne%
\ \protect \BOthers {.}}{%
{\protect \APACyear {2007}}%
}]{%
osborne2007non}
\APACinsertmetastar {%
osborne2007non}%
\begin{APACrefauthors}%
Osborne, P\BPBI E.%
, Foody, G\BPBI M.%
\BCBL {}\ \BBA {} Su{\'a}rez-Seoane, S.%
\end{APACrefauthors}%
\unskip\
\newblock
\APACrefYearMonthDay{2007}{}{}.
\newblock
{\BBOQ}\APACrefatitle {Non-stationarity and local approaches to modelling the
  distributions of wildlife} {Non-stationarity and local approaches to
  modelling the distributions of wildlife}.{\BBCQ}
\newblock
\APACjournalVolNumPages{Diversity and Distributions}{13}{3}{313--323}.
\PrintBackRefs{\CurrentBib}

\bibitem [\protect \citeauthoryear {%
{R Core Team}%
}{%
{R Core Team}%
}{%
{\protect \APACyear {2019}}%
}]{%
R}
\APACinsertmetastar {%
R}%
\begin{APACrefauthors}%
{R Core Team}.%
\end{APACrefauthors}%
\unskip\
\newblock
\APACrefYearMonthDay{2019}{}{}.
\newblock
{\BBOQ}\APACrefatitle {R: A Language and Environment for Statistical Computing}
  {R: A language and environment for statistical computing}{\BBCQ}\
  [\bibcomputersoftwaremanual].
\newblock
\APACaddressPublisher{Vienna, Austria}{}.
\newblock
\begin{APACrefURL} \url{https://www.R-project.org/} \end{APACrefURL}
\PrintBackRefs{\CurrentBib}

\bibitem [\protect \citeauthoryear {%
Siino%
, Adelfio%
, Mateu%
, Chiodi%
\BCBL {}\ \BBA {} D’alessandro%
}{%
Siino%
\ \protect \BOthers {.}}{%
{\protect \APACyear {2017}}%
}]{%
siino2017spatial}
\APACinsertmetastar {%
siino2017spatial}%
\begin{APACrefauthors}%
Siino, M.%
, Adelfio, G.%
, Mateu, J.%
, Chiodi, M.%
\BCBL {}\ \BBA {} D’alessandro, A.%
\end{APACrefauthors}%
\unskip\
\newblock
\APACrefYearMonthDay{2017}{}{}.
\newblock
{\BBOQ}\APACrefatitle {Spatial pattern analysis using hybrid models: an
  application to the Hellenic seismicity} {Spatial pattern analysis using
  hybrid models: an application to the hellenic seismicity}.{\BBCQ}
\newblock
\APACjournalVolNumPages{Stochastic Environmental Research and Risk
  Assessment}{31}{}{1633--1648}.
\PrintBackRefs{\CurrentBib}

\bibitem [\protect \citeauthoryear {%
Siino%
, D'Alessandro%
, Adelfio%
, Scudero%
\BCBL {}\ \BBA {} Chiodi%
}{%
Siino%
\ \protect \BOthers {.}}{%
{\protect \APACyear {2018}}%
}]{%
siino2018multiscale}
\APACinsertmetastar {%
siino2018multiscale}%
\begin{APACrefauthors}%
Siino, M.%
, D'Alessandro, A.%
, Adelfio, G.%
, Scudero, S.%
\BCBL {}\ \BBA {} Chiodi, M.%
\end{APACrefauthors}%
\unskip\
\newblock
\APACrefYearMonthDay{2018}{}{}.
\newblock
{\BBOQ}\APACrefatitle {Multiscale processes to describe the eastern sicily
  seismic sequences} {Multiscale processes to describe the eastern sicily
  seismic sequences}.{\BBCQ}
\newblock
\APACjournalVolNumPages{Annals of Geophysics}{}{}{}.
\PrintBackRefs{\CurrentBib}

\bibitem [\protect \citeauthoryear {%
Tarantino%
, D’Angelo%
\BCBL {}\ \BBA {} Adelfio%
}{%
Tarantino%
\ \protect \BOthers {.}}{%
{\protect \APACyear {2024}}%
}]{%
tarantino2024modelling}
\APACinsertmetastar {%
tarantino2024modelling}%
\begin{APACrefauthors}%
Tarantino, M.%
, D’Angelo, N.%
\BCBL {}\ \BBA {} Adelfio, G.%
\end{APACrefauthors}%
\unskip\
\newblock
\APACrefYearMonthDay{2024}{}{}.
\newblock
{\BBOQ}\APACrefatitle {Modelling Three-Dimensional Point Patterns} {Modelling
  three-dimensional point patterns}.{\BBCQ}
\newblock
\BIn{} \APACrefbtitle {Scientific Meeting of the Italian Statistical Society}
  {Scientific meeting of the italian statistical society}\ (\BPGS\ 588--594).
\PrintBackRefs{\CurrentBib}

\bibitem [\protect \citeauthoryear {%
Ulm%
}{%
Ulm%
}{%
{\protect \APACyear {1991}}%
}]{%
ulm1991statistical}
\APACinsertmetastar {%
ulm1991statistical}%
\begin{APACrefauthors}%
Ulm, K.%
\end{APACrefauthors}%
\unskip\
\newblock
\APACrefYearMonthDay{1991}{}{}.
\newblock
{\BBOQ}\APACrefatitle {A statistical method for assessing a threshold in
  epidemiological studies} {A statistical method for assessing a threshold in
  epidemiological studies}.{\BBCQ}
\newblock
\APACjournalVolNumPages{Statistics in Medicine}{10}{3}{341--349}.
\PrintBackRefs{\CurrentBib}

\end{thebibliography}

\end{document}